\documentclass[twocolumn, trackchanges]{aastex7}

\newcommand{\mjysr}{MJy\,sr$^{-1}$}
\newcommand{\um}{\,\micron}
\newcommand{\hii}{\textsc{Hii}}
\newcommand{\Spitzer}{\textit{Spitzer}}

\usepackage{amsmath}

\newcommand\joneval{0.1} 
\newcommand\jtwoval{2} 
\newcommand\percentMapping{27} 
\newcommand\percentLowres{67} 
\newcommand\wiseThreeFWHM{12} 
\newcommand\shardLengthSL{10.44} 
\newcommand\shardLengthLL{32.25} 
\newcommand\nShards{6,213,570} 
\newcommand\zodiModelPercentOverestimate{10} 
\newcommand\offsetLevel{0.5} 
\newcommand\deltaZodiMax{10} 
\newcommand\targDepth{50} 

\usepackage{float}

\newcommand{\ritter}{Ritter Astrophysical Research Center, Department of Physics \& Astronomy, University of Toledo, Toledo, OH 43606, USA}

\begin{document}

\title{SIMLA: The \Spitzer\ Infrared Spectrograph Mapping Legacy Archive}

\author[0009-0001-6065-0414]{Grant P. Donnelly}
\affiliation{\ritter}
\email[show]{Grant.Donnelly@rockets.utoledo.edu}

\author[0000-0003-2093-4452]{Cory M. Whitcomb}
\affiliation{\ritter}
\email{Cory.Whitcomb@rockets.utoledo.edu}

\author[0009-0005-0750-2956]{Lindsey Hands}
\affiliation{Department of Astronomy \& Astrophysics, University of California, San Diego,\\ 9500 Gilman Drive, La Jolla, CA 92093, USA}
\email{lhands@ucsd.edu}

\author[0000-0003-0014-0508]{Sara E. Duval}
\affiliation{\ritter}
\email{Sara.Duval@rockets.utoledo.edu}

\author[0000-0003-1545-5078]{J.-D. T. Smith}
\affiliation{\ritter}
\email{JD.Smith@utoledo.edu}

\author[0000-0002-4378-8534]{Karin Sandstrom}
\affiliation{Department of Astronomy \& Astrophysics, University of California, San Diego,\\ 9500 Gilman Drive, La Jolla, CA 92093, USA}
\email{kmsandstrom@ucsd.edu}

\author[]{David Carroll}
\affiliation{\ritter}
\email{dacarroll2000@gmail.com}

\author[0009-0002-7917-2384]{McKenna Dowd}
\affiliation{Department of Physics, University of Texas at Arlington, Arlington, TX 76019, USA}
\email{mckenna.dowd@uta.edu}

\author[0000-0001-7449-4638]{Brandon S. Hensley}
\affiliation{Jet Propulsion Laboratory, California Institute of Technology, 4800 Oak Grove Drive, Pasadena, CA 91109, USA}
\email{brandon.s.hensley@jpl.nasa.gov}

\author[0000-0001-9162-2371]{Leslie K. Hunt}
\affiliation{INAF—Osservatorio Astrofisico di Arcetri, Largo E. Fermi 5, 50125 Firenze, Italy}
\email{leslie.hunt@inaf.it}

\author[]{Edward Walsh}
\affiliation{Homer L. Dodge Department of Physics and Astronomy, University of Oklahoma, Norman, OK 73019, USA}
\email{edward.andrew.walsh@gmail.com}

\author[]{Julie Watson}
\affiliation{\ritter}
\email{juliewatsonaw@gmail.com}

\correspondingauthor{Grant P. Donnelly}

\begin{abstract}

We present the \Spitzer/IRS Mapping Legacy Archive (SIMLA); a complete set of mid-infrared spectral cubes built from low-resolution mapping-mode fixed-target observations from \Spitzer/IRS (5.2--38\,\micron, R$\sim$60--130). Contained in this dataset are spectral maps for several hundred spatially-resolved and unresolved objects, including galaxies, molecular clouds, supernova remnants, \hii\ regions, and more. Each cube has been carefully treated to remove astronomical foregrounds and backgrounds as well as detector effects using a novel pipeline. Cube assembly was facilitated by the \texttt{CUBISM} code, which included automatic detection and removal of bad pixels. We describe the SIMLA pipeline for reducing and validating the cubes, and we show that synthetic photometry derived from SIMLA spectra and corresponding WISE photometry typically agree within a few percent. SIMLA products and documentation related to their use will soon be available at the NASA/IPAC Infrared Science Archive (DOI: \href{https://catcopy.ipac.caltech.edu/dois/doi.php?id=10.26131/IRSA655}{10.26131/IRSA655}).

\end{abstract}


\section{Introduction}\label{sec:intro}

\setcounter{footnote}{0} 

Space-based mid-infrared (MIR) spectroscopy has been pivotal for astrophysics, owing to the diminished effect of extinction and the richness of features within this regime that arise from the interstellar medium (ISM). These capabilities began with the Infrared Space Observatory \citep[ISO,][]{ISO}, followed by the Infrared Spectrograph \citep[IRS,][]{IRS} on board the \Spitzer\ Space Telescope, which led to huge gains in our understanding of the ISM within the Milky Way and other galaxies; see reviews by: \cite{genzel_2000, vanDishoeck_2004} for ISO, and see \cite{soifer_2008, armus_2020, li_2020} for \Spitzer. Now, JWST has enabled more sensitive and higher resolution observations both spatially and spectrally, but the small fields of view (FOVs) of its integral field units (IFUs) limit JWST to highly targeted observations for spectral mapping. 

\begin{figure*}
\centering
\includegraphics[width=7in]{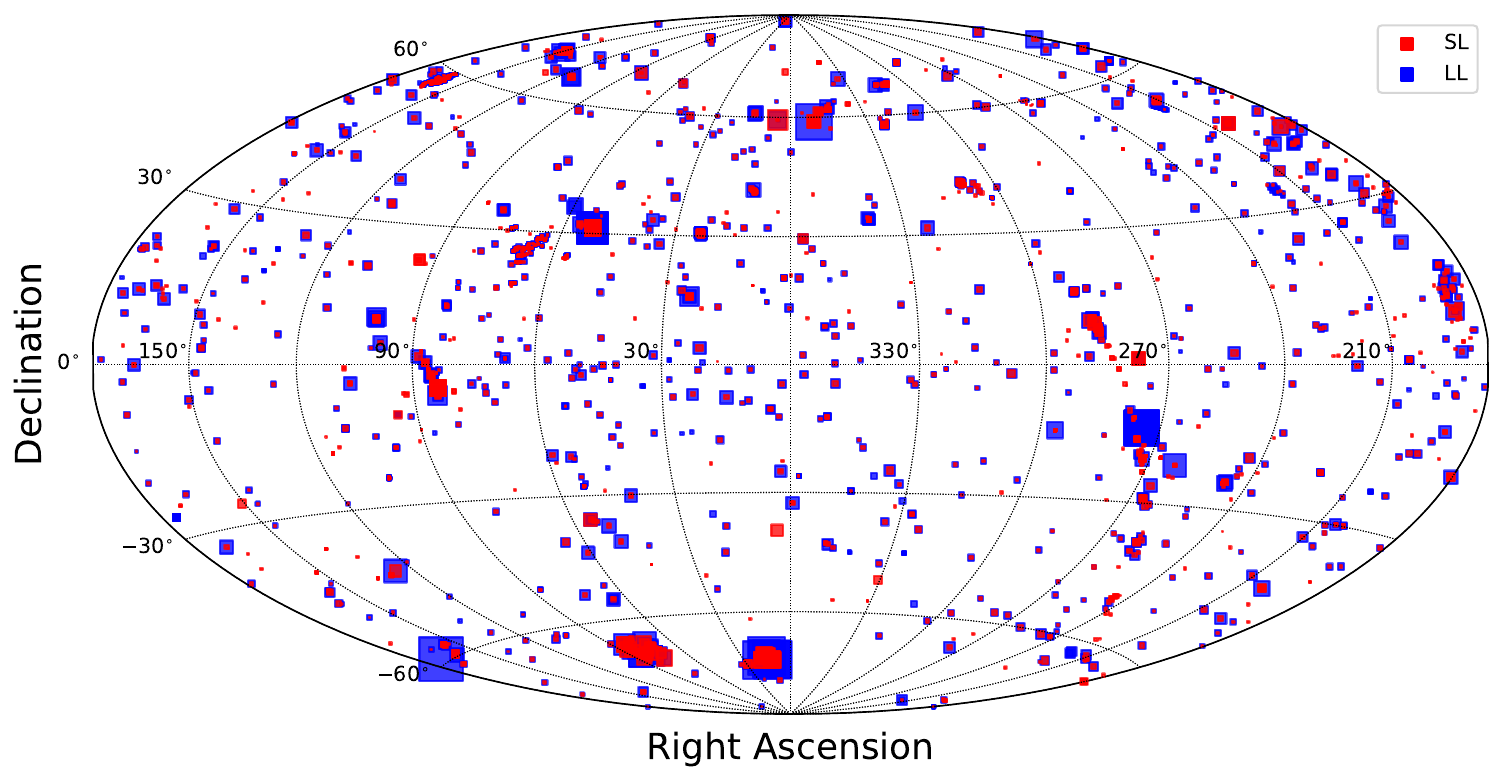}
\caption{
The distribution of SIMLA maps on-sky. Red and blue squares indicate SL and LL cubes, respectively. The size of these markers do \textit{not} represent the real fields of view, but are logarithmically scaled to the covered fields of view with an arbitrary scale factor applied to all for visibility. In this IRS sample, there is a total unique area on sky of 3.16 deg$^2$. The total unique area covered by SL is 0.94 deg$^2$ and by LL, 2.52 deg$^2$.}
\label{fig:skymap}
\end{figure*}

In contrast, the larger IRS FOVs (Table \ref{tab: irsinfo}) combined with the ability of \Spitzer\ to engage in a ``mapping mode" provided a wealth of over 5000 hours of MIR spectral observations, the spatial extent of which cannot be reproduced by any existing or currently planned facility. The totality of \Spitzer/IRS maps cover over three square degrees of sky, a larger area of the sky than any other MIR spectrometer (see Figure~\ref{fig:skymap}). The IRS was sensitive between 5.2 and 38\,\micron, covering nearly all of the major emission features from polycyclic aromatic hydrocarbons, as well as $\mathrm{H_2}$ rotational lines and a multitude of atomic and ionized gas lines. The spectral coverage is comparable to JWST/MIRI-MRS, which is sensitive to 4.9 - 27.9\,\micron, but the IRS allows for observations of the 28\,\micron\ $\mathrm{H_2}$ rotational line. The value of IRS maps is arguably enhanced even further in the era of JWST, given their ability to provide context for high resolution zoom-ins and to identify important targets for new observations.

Mapping-mode IRS observations were performed by stepping the slit across targets and then assembling the data into 3-dimensional (two spatial, one spectral) data ``cubes" that are essentially equivalent to data produced using IFUs \citep[see][]{cubism_paper, IRS}. Similar  slit-stepping techniques continue to serve as a wider-field alternative to IFU spectroscopy both from the ground \citep[e.g.,][]{TYPHOON_slitstep} and from space \citep[e.g.,][]{MSA-3D}. Although the last IRS observation concluded in 2009, there is still no repository of ready-to-use IRS cubes, and a significant fraction of the existing observations remain unpublished in part because users must manually produce these cubes individually using specialized software \citep[\texttt{CUBISM,}][]{cubism_paper}. In this work, we present the \Spitzer/IRS Mapping Legacy Archive (SIMLA), which delivers uniformly reduced spectral cubes for nearly\footnote{Except for observations of moving objects, i.e., in the Solar System, which require additional specialized map processing, and for IRS calibration data.} all low spectral resolution IRS maps from the \Spitzer\ archive.

Backgrounds are a particularly important part of IRS data reduction for two main reasons. First, many otherwise functional pixels on the IRS detector had nonzero offsets called ``pedestals" \citep[see][]{irs_handbook} that were stable over a poorly defined timescale. However, the IRS did not have a shutter to facilitate dark calibration images, so background observations that are nearby in time were critical to mitigate the pixel stochasticity as well as subtract out truly nonfunctional pixels. Second, scattered and emitted light from Solar System dust (i.e., the zodiacal light) creates a significant or dominant foreground signal in MIR spectra. Zodiacal light models are useful for removing the bulk of this emission, but sky background observations are necessary to account for small variations in the spectral shape and intensity levels to study low-surface brightness objects. Many observing programs did not include any background or only obtained sparse backgrounds that could be a dominant contributor to noise. Thus, the SIMLA pipeline is primarily concerned with the creation of tailored backgrounds for each cube. 

\begin{figure*}
\centering
\includegraphics[width=7in]{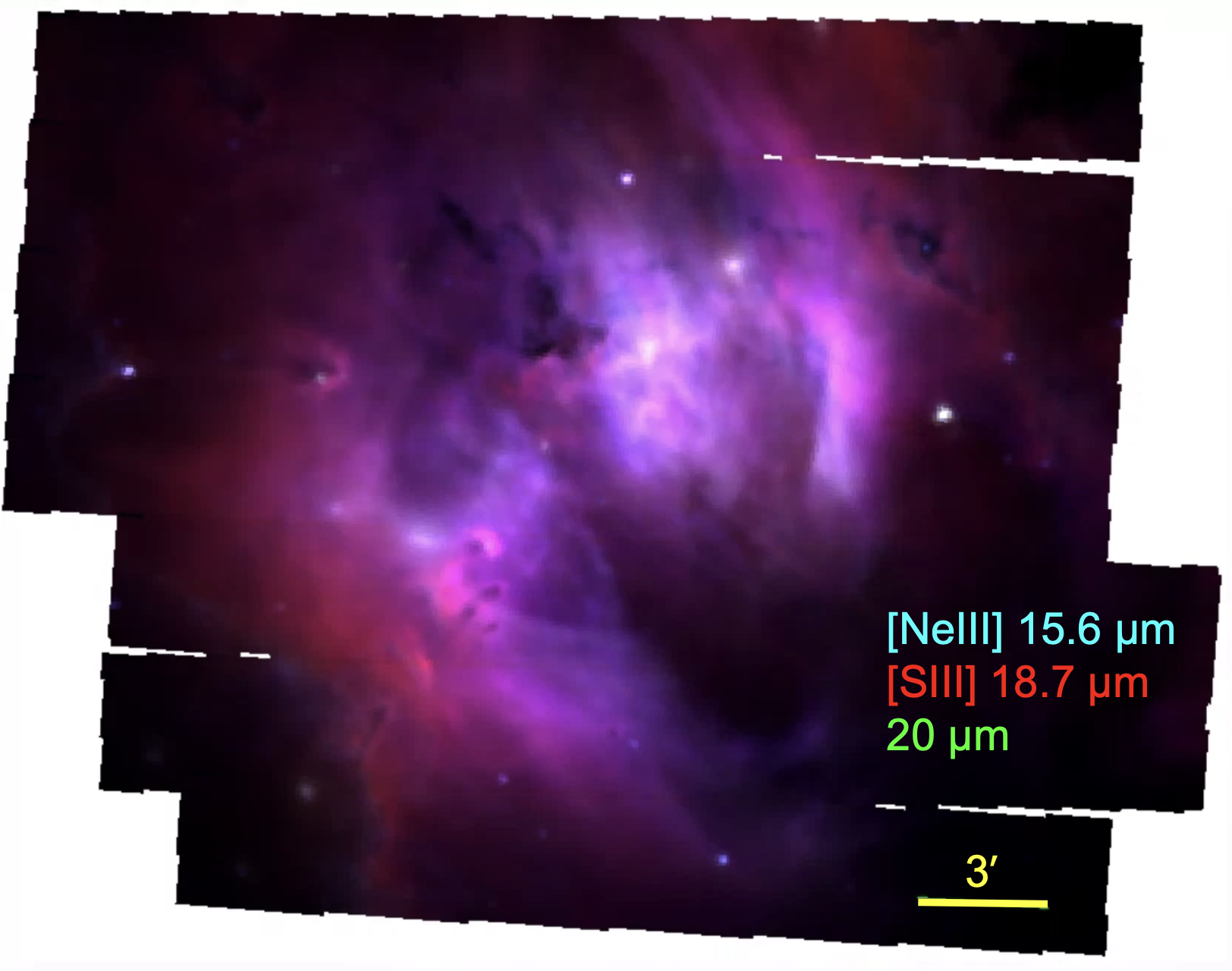}
\includegraphics[width=3.5in]{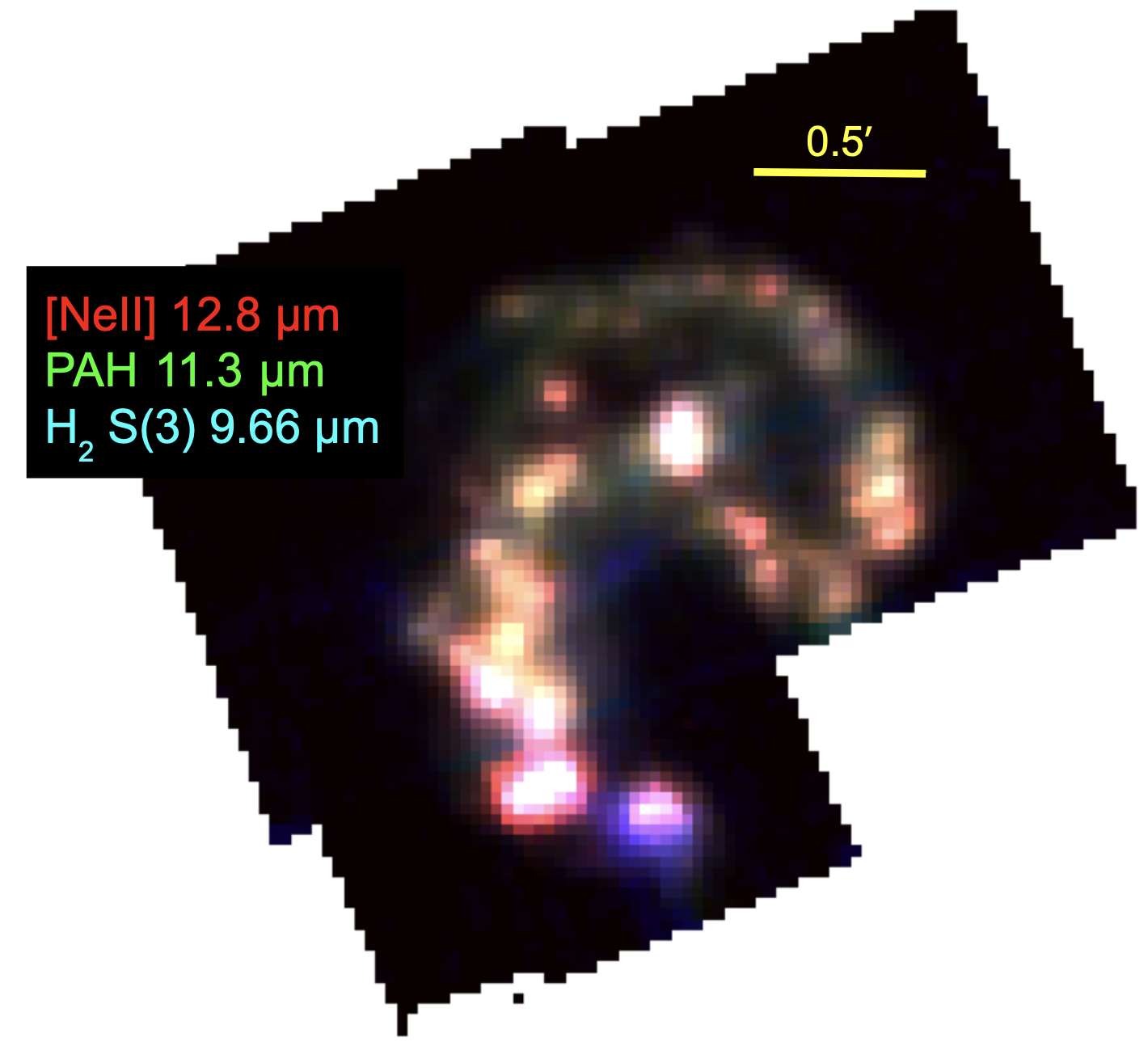}
\includegraphics[width=3.5in]{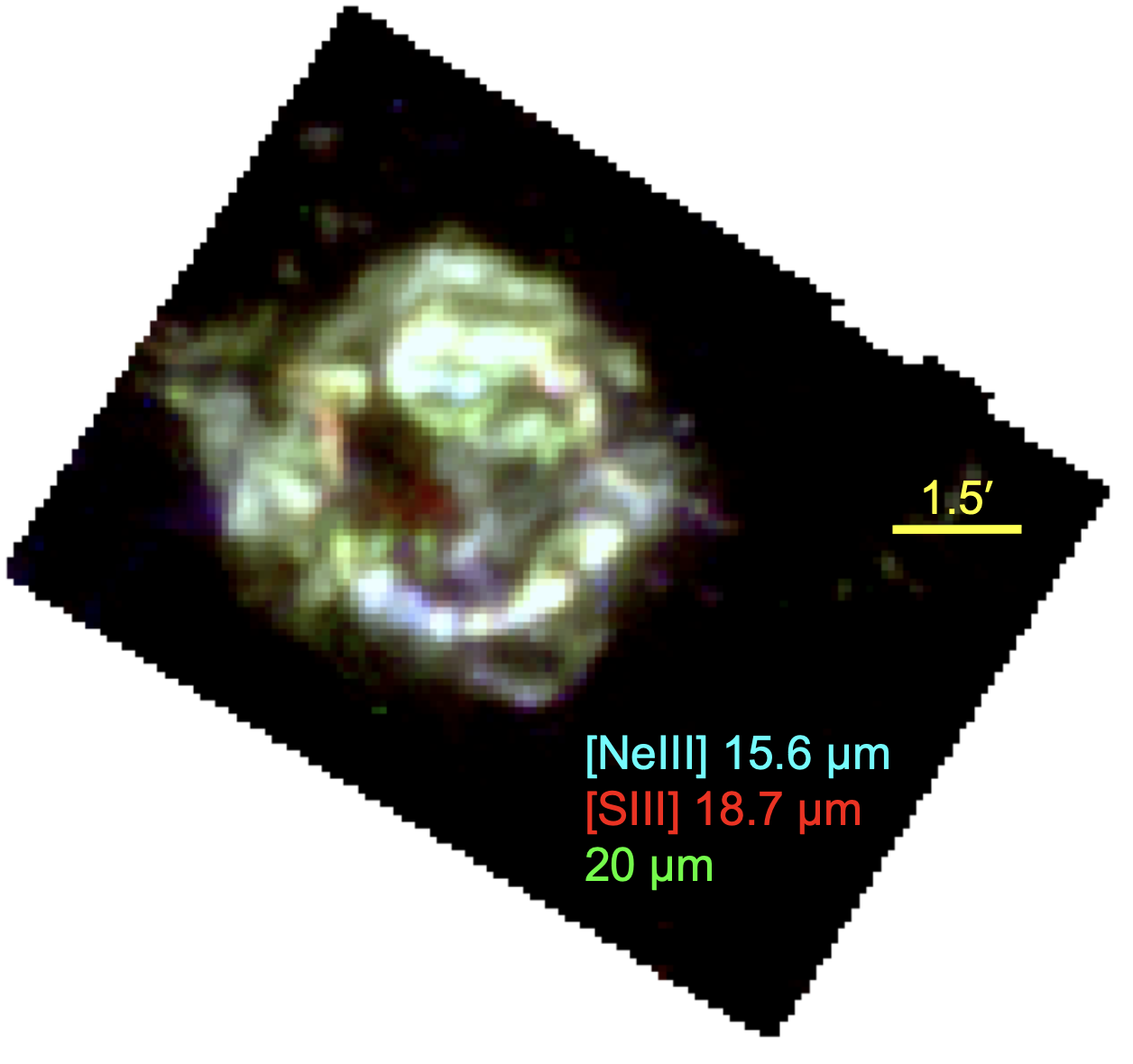}
\caption{
Three-color images made using mosaics of SIMLA spectral cubes of the Eagle Nebula (top), the Antennae galaxies (bottom left), and the Cassiopeia A supernova remnant (bottom right). For each image, the pixel values of each color correspond to the slice of a SIMLA cube mosaic at the listed wavelength.
}
\label{fig:prettypics}
\end{figure*}

\begin{figure*}
\centering
\includegraphics[width=7in]{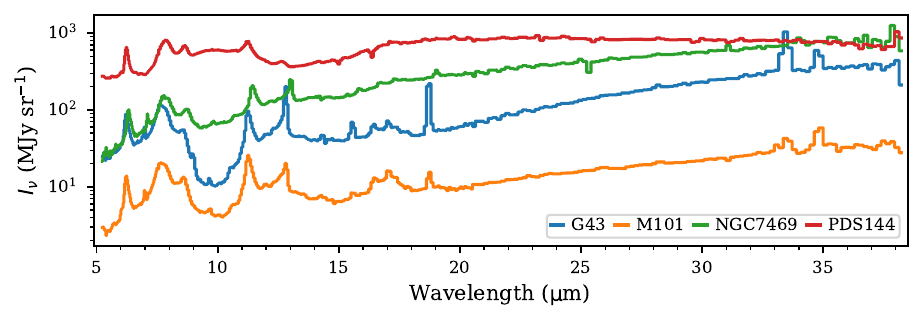}
\caption{
Example spectra from SIMLA spectral cubes of the Galactic \hii\ region G43 (blue), the star-forming galaxy M101 (orange), the luminous infrared galaxy NGC\,7469 (green), and the binary Herbig Ae stars PDS144 (red).
}
\label{fig:examplespecs}
\end{figure*}

The goals of SIMLA are to leverage the entire IRS archive to: 1) build a nearly complete set of data cubes for \Spitzer/IRS mapping mode observations; 2) create and use custom backgrounds for all cubes, dramatically improving data for programs that did not collect backgrounds or obtained poor backgrounds; 3) generate sufficiently deep backgrounds to achieve the best possible signal-to-noise ratio (S/N) in each mapping observation; 4) remove the zodiacal light foreground using a combined model/data approach and; 5) provide the built background-subtracted cubes to the community. In this paper, we describe the first SIMLA data release of cubes from the low spectral resolution modules (R$\sim60-130$), which make up a majority of mapping mode observations ($\sim\percentLowres\%$), and $\sim\percentMapping\%$ of all existing IRS data (e.g., Figure~\ref{fig:prettypics} for three-color mosaics from SIMLA cubes, Figure~\ref{fig:examplespecs} for example spectra from SIMLA cubes). 

This work is organized as follows: in Section~\ref{sec:datamodels}, we briefly describe the data and models that we used. The complete details of the SIMLA pipeline are described in Section~\ref{sec:pipeline}, and data validations are described in Section~\ref{sec:QA}. In Section~\ref{sec:caveats}, we provide additional details about SIMLA cubes, such as noise statistics and sensitivity limitations, that are useful for users to know. Additional details about SIMLA data products are described in the data delivery document provided alongside the archive.


\section{Data and Models}\label{sec:datamodels}

In this section, we summarize the details of the data, instruments, and software that we use for the background creation and the cube assembly. This includes the IRS data itself (Section~\ref{sec:spitzer_IRS}) and the CUBISM software (Section~\ref{sec:CUBISM}), as well as observations from the Wide-field Infrared Explorer \citep[WISE, Section~\ref{sec:WISE};][]{WISE} and model of the MIR zodiacal foreground from \cite{kelsall_1998}, which were used in the creation of backgrounds for SIMLA cubes (Section~\ref{sec:zodimodel}). 

\subsection{Spitzer/IRS}\label{sec:spitzer_IRS}

Further details and complete explanations of all topics in this subsection can be found in the IRS Instrument Handbook \citep{irs_handbook}. Certain basic characteristics of \Spitzer/IRS are summarized in Table~\ref{tab: irsinfo}.

\begin{figure*}
\centering
\includegraphics[width=7in]{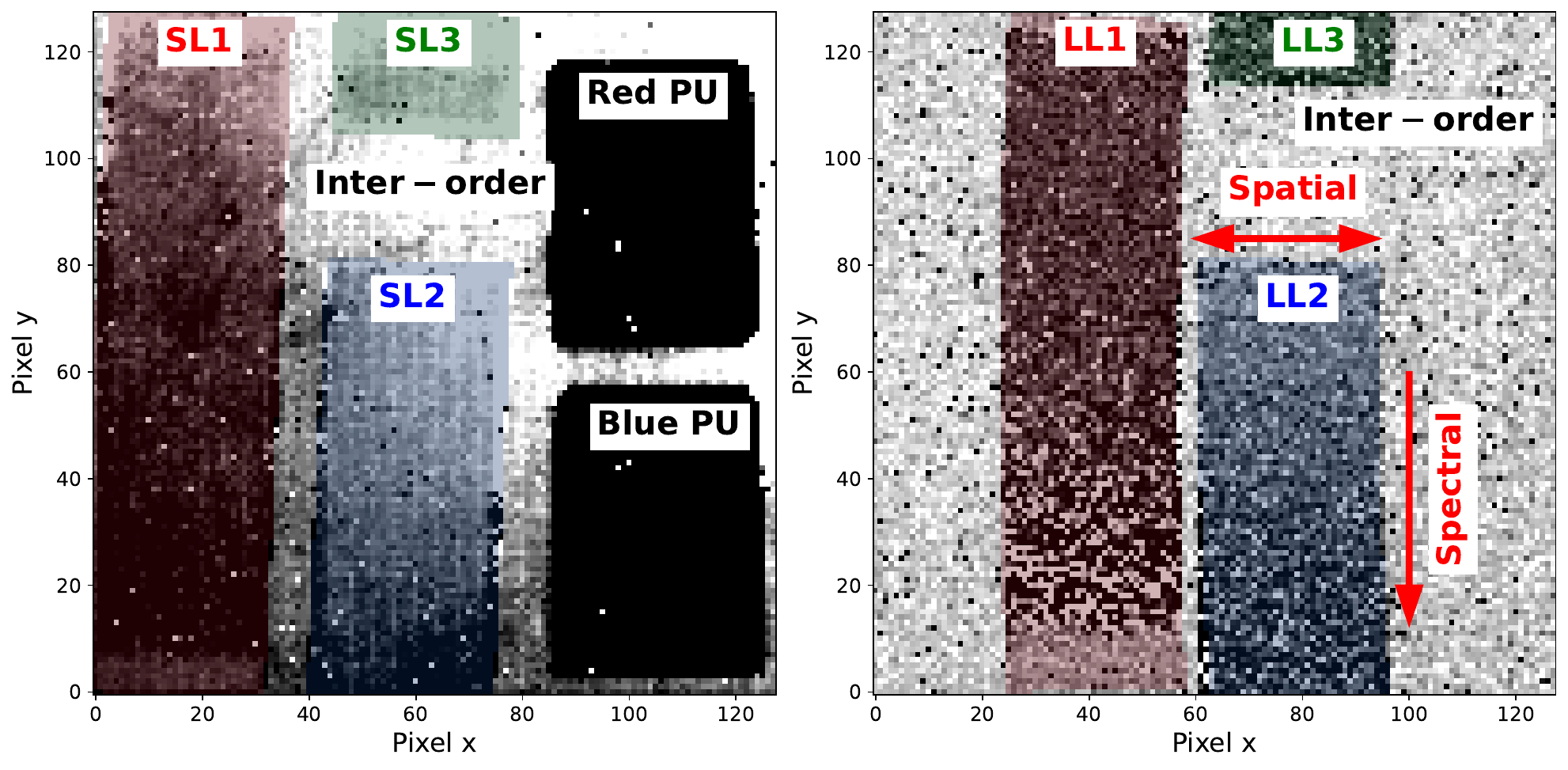}
\caption{
Basic calibrated data (BCD) example images for \Spitzer/IRS with major sections labeled. The left and right panels show BCDs for the short-low (SL) and long-low (LL) modules, respectively. For both, the red and blue sections highlight the \texttt{wavesamp} regions for the sub-slits within either module. The green sections are the ``bonus orders" (SL3, LL3), on which light from the 2nd slit (SL2, LL2) is sampled within a wavelength range overlapping the red and blue ranges. BCDs for SL contain the light from the peak-up (PU) photometer arrays. Both of these BCDs are from pointings containing no bright sources and they have minimal zodiacal emission intensity. Note the spatially and spectrally-varying artifact in the SL BCD that persists throughout the spectral orders as well as into the inter-order region.
}
\label{fig:irs_diagram}
\end{figure*}

\subsubsection{IRS Slits}\label{sec:IRS_apertures}

Spitzer/IRS had both ``low'' (R$\sim60-130$) and ``high" (R$\sim600$) resolution modes, and this SIMLA release focuses on the low-resolution data. On the \Spitzer\ focal plane, two long-slit apertures were used to collect low-resolution spectra, each corresponding to its own wavelength range with a slight overlap. Each of these slits were divided into two inline sub-slits; two sub-slits make up the short-wavelength low-resolution channel (short-low or SL), SL1 (7.5 -- 14.7\,\micron) and SL2 (5.2 -- 7.6\,\micron), and two larger sub-slits make up the long-wavelength low-resolution channel (long-low or LL), LL1 (20.6 -- 38\,\micron) and LL2 (14.3 -- 21.1\,\micron). The data from a pair of sub-slits were always obtained simultaneously, but data from the separate SL and LL modules could not be. Additionally, both the SL and LL detectors feature an overlapping ``bonus order" (SL3 and LL3, 7.4 -- 8.7\,\micron\ and 19.9 -- 21.1\,\micron, respectively) which in each case use the aperture of the second order sub-slit (SL2 or LL2).

The simultaneous but offset observing in the sub-slits produced a large number of ``outrigger" background observations when one sub-slit was observing a compact or point source and the off-source sub-slit was observing the source-free sky. This setup was often used strategically in order to collect background observations during integration on-source. However, if an extended target source filled the whole FOV of both slits, a separate pointing would have been required as a dedicated background. Observations from both the outrigger and dedicated pointings are used for SIMLA backgrounds, as well as any circumstantially dark parts of observations (see Section~\ref{sec:shards}).

\subsubsection{Area Observation Requests and Basic Calibrated Data}\label{sec:AORs_and_BCDs}

Groups of IRS observations over a single area are related by their area observation request (AOR). While some astronomical objects were covered by multiple AORs, these may have used different observational settings (for example, integration times) or were separated by a significant amount of time, meaning that the noise characteristics between the observations in different AORs are distinct. We generally consider individual mapping-mode AORs as the basic unit of what is made into a SIMLA cube. Additional processing to stitch together cubes from multiple AORs is not included in the SIMLA delivery, but can be accomplished by users with existing mosaicking tools\footnote{Such as the \texttt{reproject} package (\url{https://github.com/astropy/reproject}) or Montage (\url{http://montage.ipac.caltech.edu/}).}.

IRS spectral images for a single pointing within an AOR are called basic calibrated data (BCDs). These images, along with their associated uncertainty images derived primarily from nondestructive detector ramp readout, were generated by the Spitzer Science Center (SSC) BCD pipeline.  SIMLA cubes are built from the SSC IRS pipeline version S18.18.0. See Figure~\ref{fig:irs_diagram} for examples of both a SL and a LL BCD. For all IRS observations and for each module (e.g., SL or LL), observers could specify one of four choices for the \texttt{RAMPTIME}, which is the duration of a data collection event. An observer-specified number of these events sums to the total integration time associated with a BCD. Because different values of the \texttt{RAMPTIME} have different noise characteristics, we treat BCDs with different \texttt{RAMPTIME} values separately in the SIMLA pipeline. BCDs from both the SL and LL detectors contain relatively stable dark current artifacts, but they are different between each \texttt{RAMPTIME}. These patterns span the parts of the detector that are exposed to light through the slits, which we refer to as the \texttt{wavesamp} area (colored sections in Figure~\ref{fig:irs_diagram}), as well as the ostensibly dark region of the detector which we refer to as the inter-order (IO) region. In the left panel of Figure~\ref{fig:irs_diagram}, the dark current pattern appears as a pixel value excess emanating from the bottom-left corner (see Section~\ref{sec:superdarks}). 

\subsubsection{Observing Modes}\label{sec:observing_modes}

The IRS could be used in two primary modes: staring-mode and mapping-mode. There is no difference at the BCD level between either observing mode, meaning that staring-mode BCDs may be used in the constructed backgrounds for mapping-mode cubes, and vice-versa. Observations in the staring-mode were typically performed using two pointings separated by 1/3 of the slit aperture length, called ``nod positions." This was done to mitigate the effects of cosmic rays and ensure pixel redundancy. The staring mode was often used for point sources, and constitutes the larger fraction of IRS data. Because of the simultaneous observing in the SL1/SL2 or LL1/LL2 sub-slits, while staring mode observations towards a compact target were occurring in one sub-slit, the other was often observing nearby dark sky. This makes the staring mode data a rich source for constructing backgrounds for the mapping mode data. Indeed, recent work by \cite{boersma_2024} has used the large number of staring mode spectra pointed at sky to study Galactic diffuse ISM dust emission. 

Mapping-mode observations were made by stepping the slits in the parallel and perpendicular directions across a target with a spacing determined by the observer; this mode is the focus of SIMLA. A single mapping AOR could cover one area or be split up into clusters of maps. For the clusters, the SIMLA pipeline produces separate maps for each pointing area. Some programs were performed in the mapping-mode, but were set up with no perpendicular and/or parallel steps, and so are similar to staring mode observations. However, since these programs are tagged as mapping-mode, they are not treated as different by the SIMLA pipeline and they are also produced into 3D spectral cubes\footnote{The SIMLA pipeline uses \texttt{CUBISM} settings designed for mapping-mode observations, such as applying a slit-loss correction factor, that is not tailored for staring-mode observations. Absolute flux values for cubes with no perpendicular slit steps may be affected.}. In addition to determining how wide of an area is covered by a map, the step size for the slit can also serve to promote pixel redundancy in the same way that the nodding does for staring-mode. Fully-sampled maps have a step size that is at most one-half of the slit size, but many maps opted for a sparser sampling in order to cover a wider area.

\begin{deluxetable}{lcccc}
\tabletypesize{\scriptsize}
\tablehead{
\colhead{\textbf{Suborder}} &
\colhead{Slit Size} &
\colhead{Wavelength} &
\colhead{Pixel Scale} &
\colhead{Resolving} \\ [-3mm]
\colhead{ } &
\colhead{[\arcsec $\times$ \arcsec]} &
\colhead{Range [\micron]} &
\colhead{[\arcsec/pixel]} &
\colhead{Power [$R$]}
}
\startdata
SL2 & 57 $\times$ 3.6 & 5.24 - 7.60 & 1.8 & 60 - 72 \\
SL3 & ... & 7.37 - 8.67 & 1.8 & 60 - 72 \\
SL1 & 57 $\times$ 3.7 & 7.53 - 14.74 & 1.8 & 72 - 127 \\
LL2 & 168 $\times$ 10.5 & 14.27 - 21.05 & 5.1 & 57 - 63 \\
LL3 & ... & 19.91 - 21.10 & 5.1 & 57 - 63 \\
LL1 & 168 $\times$ 10.7 & 20.56 - 38.42 & 5.1 & 63 - 126 \\
SH & 11.3 $\times$ 4.7 & 9.97 - 19.44 & 2.3 & 600 \\
LH & 22.3 $\times$ 11.1 & 19.12 - 37.16 & 4.5 & 600 \\
\enddata
\caption{Summary of Spitzer/IRS characteristics. Slit size dimensions are not given for the SL3 and LL3 suborders as these share the SL2 and LL2 sub-slits, respectively. \label{tab: irsinfo}}
\end{deluxetable}

\subsubsection{LL Bias Change}\label{sec:LL_bias}

In order to combat the increasing number of bad pixels in LL images in the latter portion of the IRS mission, the bias and temperature were changed for the LL module beginning with IRS Campaign 45 ($\mathrm{MJD}>54403$). These changes significantly affected the noise characteristics and sensitivity of the detector. For this reason, we always treat LL observations on either side of the bias change as if they were entirely separate channels for background construction.

\subsection{Zodiacal Emission Model}\label{sec:zodimodel}

The 5.2 -- 38\,\micron\ range covered by \Spitzer/IRS is contaminated by significant foreground emission that arises from warm dust in the Solar System; this emission is often referred to as zodiacal light \citep[e.g.,][]{reach_2003}. The zodiacal light is time variable, as observers within the Solar System view it from different locations within the spatial distribution of dust throughout an orbit around the Sun. Across all IRS observations its intensity can vary by almost a factor of five, but the spectral shape does not vary significantly. In this work, we used the model from \cite{kelsall_1998}, which estimates various components of foreground and background emission. We primarily use the zodiacal emission model (ZEM), evaluated for the ephemeris of \Spitzer, but we also use the Galactic ISM estimates as a cut for finding suitably dark background regions.

\subsection{WISE}\label{sec:WISE}

We used all-sky maps from the Wide-Field Infrared Survey Explorer \citep[WISE,][]{WISE, WISE2} to identify source-free (``dark") IRS apertures to use as components in backgrounds. Later, we also use the WISE maps to validate the flux values of the final SIMLA cubes (Section~\ref{sec:WISE_comparison}). The W3 band centered at 12\,\micron\ ($\sim8.5-17\,\micron$) is ideal for these purposes because of its sensitivity in addition to its overlap with both low-resolution IRS channels. We retrieved all 5,076 WISE W3 images from the ALLWISE data release \citep[][]{WISEDOI} that have a sky footprint containing IRS apertures from the Infrared Science Archive (IRSA); these images have been smoothed to the \wiseThreeFWHM\,\arcsec\ resolution of the W4 band (22\,\micron). We converted each image to physical units of \mjysr\ by following the documentation\footnote{\url{https://irsa.ipac.caltech.edu/data/WISE/docs/release/All-Sky/expsup/sec2_3f.html}}. Because they are not background subtracted, we applied our own local background subtraction for each W3 image (described in Section~\ref{sec:judge1}) in order to locate IRS observations on source-free sky and to compare flux levels with background-subtracted SIMLA cubes. 

\subsection{CUBISM}\label{sec:CUBISM}

To create the spectral data cubes out of the BCDs in an IRS mapping AOR, we use the CUBISM code\footnote{\url{https://irsa.ipac.caltech.edu/data/SPITZER/docs/dataanalysistools/tools/cubism/}}. As detailed in \citet{cubism_paper}, CUBISM is an implementation of a spectral mapping reconstruction algorithm for \Spitzer/IRS mapping observations, written in IDL.  CUBISM constructs spectral cubes fully sampling the AOR's mapping coverage (i.e., step size and map shape) by clipping the images of the slit formed at each wavelength in the detector plane of each of the four IRS spectrographs. In this way, CUBISM was a precursor to modern ``3D drizzle'' algorithms, e.g., for the JWST/MIRI-MRS image slicer \citep{Law2023}.  

The CUBISM algorithm accounts for the non-linear movement of the slit centroids as well as the varying rotation angle of the slits with respect to the detector coordinates as a function of wavelength.  At each wavelength, in each grating order, polygonal pixel fragments from each BCD contributing to the cube are projected onto and clipped within a position grid laid out on the sky, maintaining full accounting of which pixels from which input BCD images contributed to each cube pixel, and by what amount.  With good map sampling, this allows robust bad pixel flagging, utilizing the ``pixel diversity'' contributing to each position on the sky at each wavelength with custom sigma-clipping and fractional outlier thresholds. 

Low-resolution (SL and LL) orders are built into separate cubes, one for each of the low resolution grating orders 1--3 that appear on the detectors.  CUBISM provides spectro-photometric calibration applicable to extended sources using measurements of the slit loss of a point source as a function of wavelength together with the point-source based flux calibration of the default \Spitzer/IRS pipeline.  Both surface brightness and uncertainty cubes are produced, using error propagated from the input BCD and background uncertainties and associated pixel masks.

\section{SIMLA Pipeline}\label{sec:pipeline}

The SIMLA pipeline produces one cube per spectral order (SL1, SL2, etc., as applicable) for each AOR, except for AORs within the categories of \texttt{TargetMulti} and \texttt{TargetFixedCluster}, for which a separate cube per target per spectral order is made. Although multiple AORs may overlap in sky coverage, we treated each AOR separately for the purpose of building cubes. Each overlapping AOR may have been observed at very different times, may have different observational setup parameters such as \texttt{RAMPTIME}, and may have different amounts of redundancy in pixel sampling.

Most of the process in the SIMLA pipeline is oriented towards creating suitable backgrounds to subtract for each AOR, in order to remove foreground zodiacal emission and mitigate pixel stochasticity which improves the S/N of a cube. Ideally, obtaining a background would be done as a dedicated part of an observing program in the form of a separate integration that was off-source but nearby in time and in sky location. Thus, both the pixel behavior and the zodiacal emission captured by the background would be similar enough to the on-source observations to be properly subtracted out. However, many IRS observing programs did not include dedicated background pointings, and we aim to further improve those that did have background pointings by incorporating additional depth. These goals are made possible by identifying suitable dark observations from across the entire IRS mission, allowing BCDs to contribute to the backgrounds of cubes in a way that was not possible during the mission.

\begin{figure}
\centering
\includegraphics[width=0.5\textwidth]{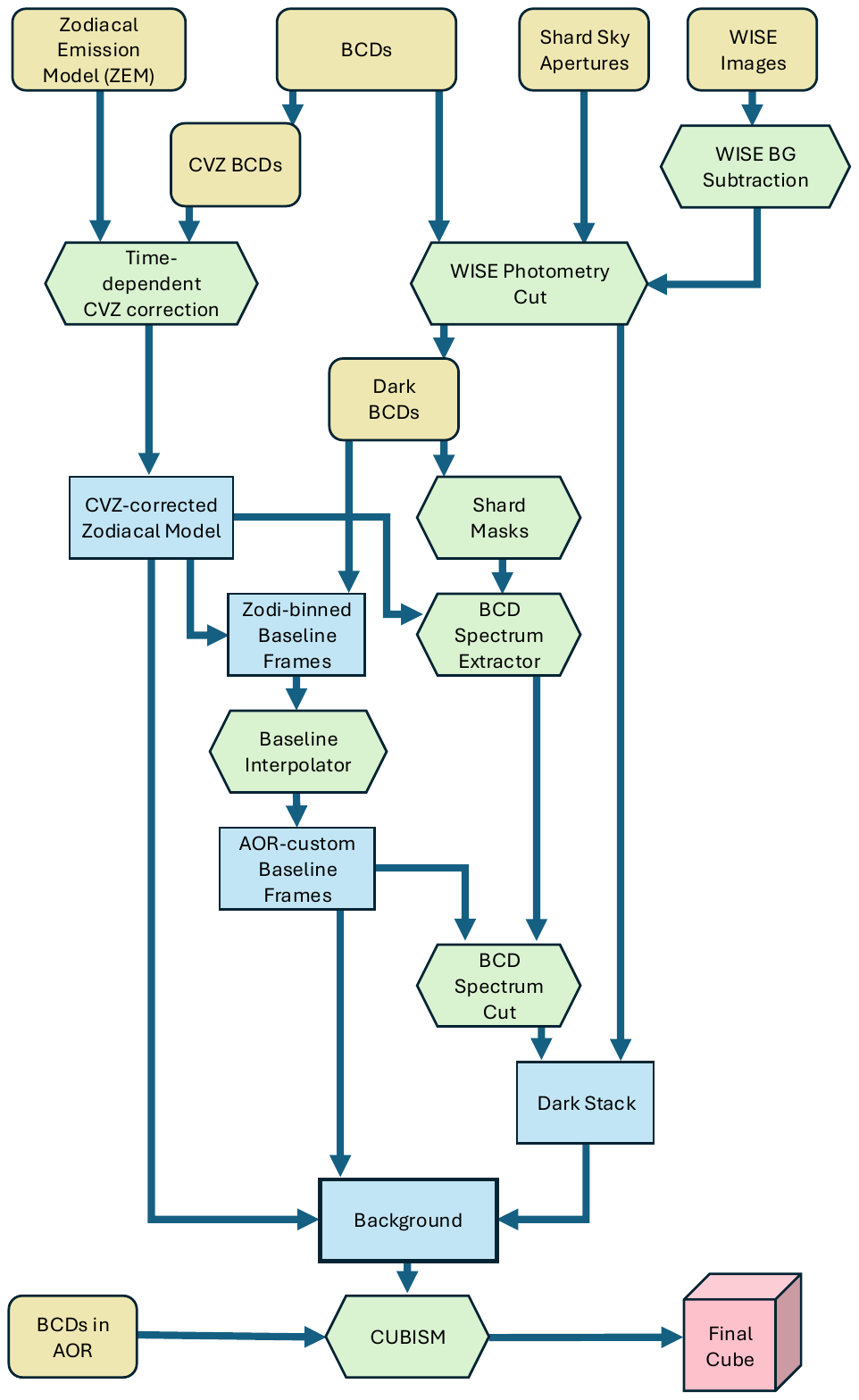}
\caption{
Schematic of the SIMLA pipeline to produce a cube. Blue boxes indicate SIMLA intermediate products derived from more basic inputs shown as yellow rounded boxes. Green hexagons are operations. Note that most of the SIMLA pipeline is oriented towards producing backgrounds.
}
\label{fig:flowchart}
\end{figure}

After the background for an AOR is built, it and the constituent BCDs of the AOR are input to \texttt{CUBISM} (Section~\ref{sec:CUBISM}) for cube building. In addition, we made use of the functionality within \texttt{CUBISM} for automatic bad pixel flagging, as described in Section~\ref{sec:using_CUBISM}. Figure~\ref{fig:flowchart} represents how each part of the pipeline, described below, fits together as a flowchart. The code for the SIMLA pipeline is available for view/use online\footnote{\url{https://github.com/simlacube/simla}}.

\subsection{Construction of Backgrounds}\label{sec:making_backgrounds}

\begin{figure}
\centering
\includegraphics[width=0.47\textwidth]{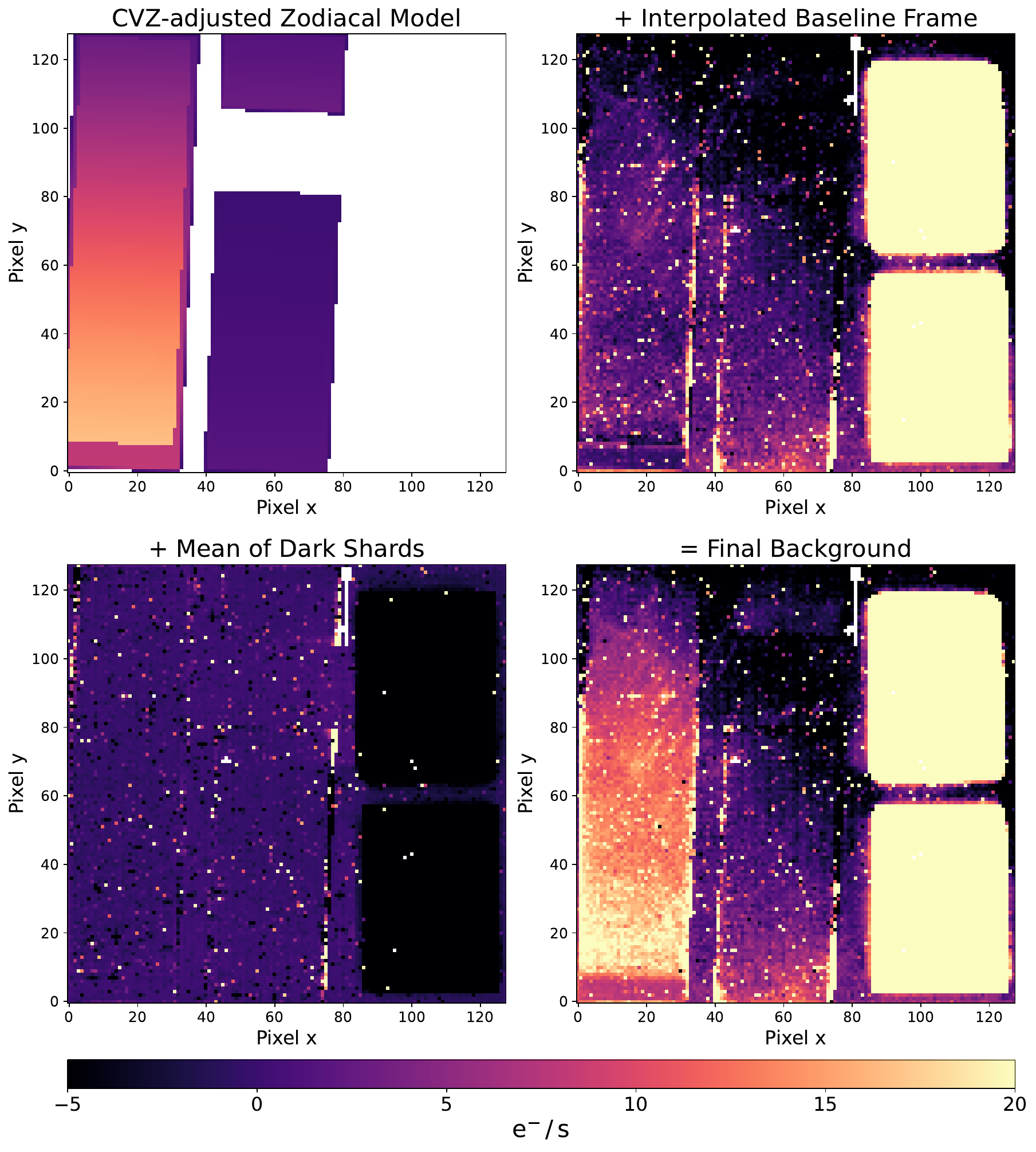}
\caption{
Examples of the three main components of SIMLA backgrounds are shown for one SL AOR: the 2D zodiacal emission model adjusted for the CVZ spectrum is in the top left (Section~\ref{sec:zodibcds}), the interpolated baseline frame scaled for the zodiacal intensity of this AOR is in the top right (Section~\ref{sec:superdarks}), and the average of the stack of dark shards is in the bottom left (Section~\ref{sec:shards}). In the bottom right panel, the sum of these three images is shown as the final background.
}
\label{fig:dark_components}
\end{figure}

 SIMLA backgrounds are composed of three major components that will be described in the following subsections. An example of each of these three components, and the final background they sum to, is shown in Figure~\ref{fig:dark_components}. They are; 1) the zodiacal light (Section~\ref{sec:zodibcds}, top-left panel in Figure~\ref{fig:dark_components}), which includes taking into account that a part of the zodiacal light has already been removed from each BCD by the use of ``super darks'' in the IRS calibration pipeline; 2) removing a persistent, static dark current feature that appears to be a property of the IRS low resolution detectors, and removing residuals correlated with the intensity of zodiacal light (Section~\ref{sec:superdarks}, top-right panel in Figure~\ref{fig:dark_components}); and 3) removing time-varying pixel pedestal offsets by averaging together a large collection of close-in-time dark observations (Section~\ref{sec:shards}, bottom-left panel in Figure~\ref{fig:dark_components}).

\subsubsection{Zodiacal Emission Models for Each AOR}\label{sec:zodibcds}

The first main component to SIMLA backgrounds deals with the zodiacal light, which requires a time-dependent correction due to the fact that the calibration super darks capture the zodiacal emission at the \Spitzer\ continuous viewing zone (CVZ), which is a sky location near the north ecliptic pole that was always accessible under the observational constraints of \Spitzer. Since the zodiacal light does not vary significantly over the duration and solid angle of any AOR, we apply this correction uniformly across an AOR, rather than on a BCD basis. For an AOR $k$, pixel $ij$ in the 2D CVZ-corrected ZEM has the value

\begin{equation} \label{eq:zodi}
\mathcal{Z}_{k, ij}(\lambda) = W_{ij}^{-1}\left\{Z_{k}(\lambda) - a(t) \hat{Z}_{\mathrm{CVZ}}(\lambda)\right\}\, ,
\end{equation}

\noindent where $Z_k(\lambda)$ is the model zodiacal spectrum, $\hat{Z}_{\mathrm{CVZ}}$ is the average ZEM spectrum at the CVZ normalized at 10.95 or 28.75\,\micron\ (for SL or LL, respectively), $a(t)$ is the amplitude of $\hat{Z}_{\mathrm{CVZ}}$ at time $t$, and $W_{ij}^{-1}$ is the inverted \texttt{wavesamp} function that projects the 1D spectrum onto a 2D BCD-like image. The details of these components are described below.

The noise within a cube generally decreases as the depth of its background increases. Thus, to assemble the background of a given cube, it is beneficial to collect off-source observations from as many AORs as possible within the time frame that the pixel offsets are coherent. To this end, it is necessary to remove the spatially and temporally varying zodiacal emission signal so that darks from different AORs contain only noise/pixel pedestals, and are therefore on equal footing to be used together in backgrounds. We used the \cite{kelsall_1998} model to generate a model zodiacal emission spectrum across the full IRS wavelength range for each AOR, taking into account the observation pointing, its time, and the ephemeris of \Spitzer. We projected these spectra onto a BCD-like 2D image by inverting the \texttt{wavesamp} function so that they can be included as part of the background for each cube.

\begin{figure*}
\centering
\includegraphics[width=7in]{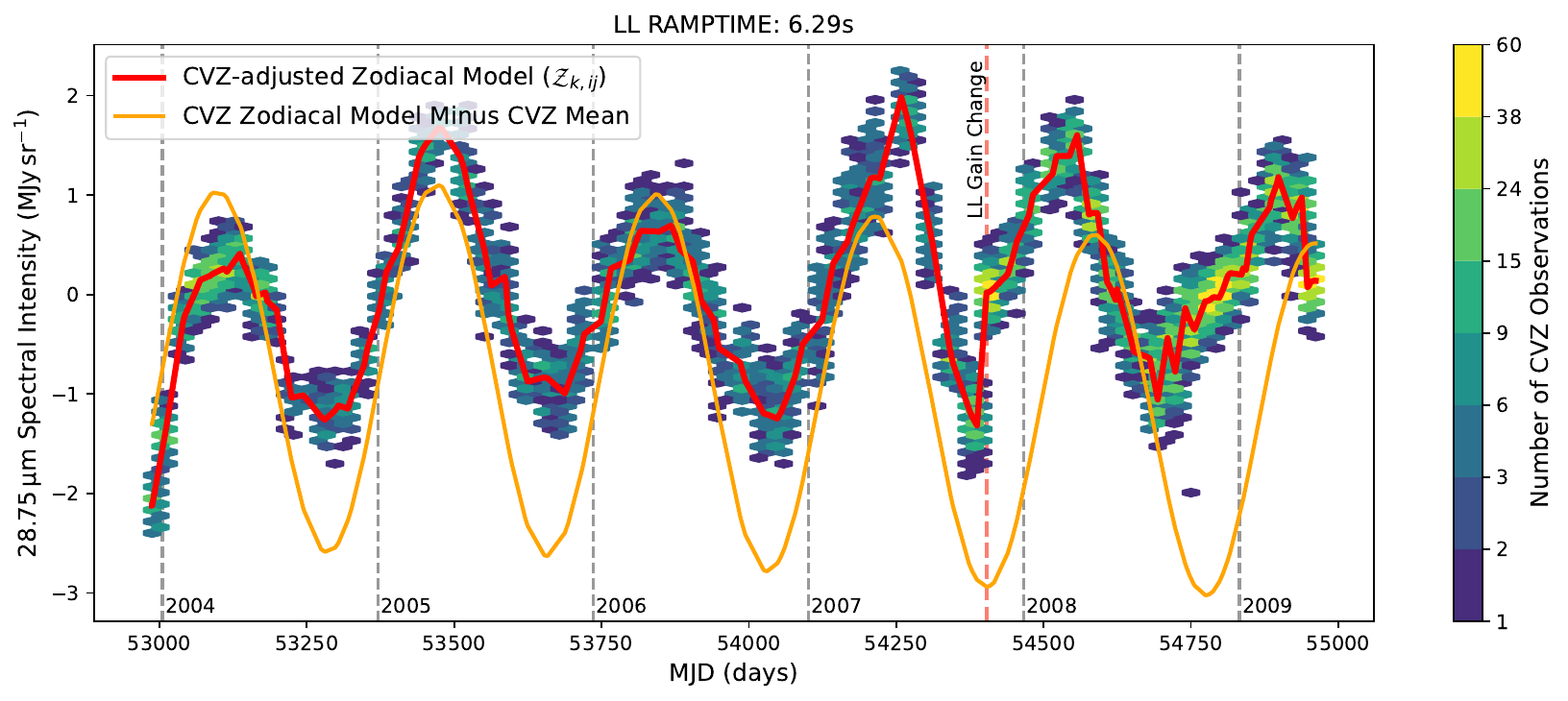}
\caption{
Observations at the continuous viewing zone (CVZ) track the time-dependent offsets in the zodiacal emission model. Here, we show these offsets as the 28.75\,\micron\ intensities (center of LL1) of observations at the CVZ for one \texttt{RAMPTIME} as colored hexagons. The orange curve shows the modeled zodiacal intensity at 28.75\,\micron\ of a given time at the CVZ minus the modeled mean intensity at the CVZ. This pattern oscillates naturally over time with the shift in viewing geometry within the non-axisymmetric zodiacal cloud. For the red curve, a scaled model of the zodiacal spectrum is subtracted instead of the mean, with the scaling factors derived from the observed data (Section~\ref{sec:zodibcds}). The offset between the red curve/colored hexagons and the orange curve results from the changing (and unknown) set of CVZ observations that were included in the creation of the super darks, capturing different mixtures of zodiacal intensity over the course of the mission.
}
\label{fig:cvztime}
\end{figure*}

Because \Spitzer/IRS lacked a shutter, it was not possible to obtain a typical ``dark'' frame for calibration. The ``super darks'' in the BCD pipeline were created by combining observations at the CVZ which have minimal (but nonzero) zodiacal light intensity. Thus, super darks contain both dark current and some level of zodiacal light. The consequence of this is that the intensities of all calibrated IRS observations are inherently relative to the zodiacal light intensity at the CVZ, and the temporal variation of the CVZ zodiacal intensity must be accounted for in the zodiacal light component of SIMLA backgrounds. 

For an observer pointing at a fixed position, the zodiacal intensity at the CVZ naturally oscillates around some average value during an orbit around the Sun (orange curve in Figure \ref{fig:cvztime}). For example, this oscillation pattern was shown by \cite{krick_2012} at the CVZ using data from \Spitzer/IRAC. Calibrated IRS observations from the SSC pipeline (i.e., BCDs) are super dark-subtracted, meaning that the zero point for IRS observations is time-dependent, reflecting this oscillation pattern. In other words, whereas the zodiacal light sets the \textit{astrophysical} minimum for the intensity of any MIR observation, the zodiacal intensity captured by a super dark sets the \textit{effective} minimum for IRS data. Because we use the ZEM as a component in SIMLA backgrounds, we must apply a correction to the model to reflect this difference.

In the SSC BCD pipeline, one super dark per \texttt{RAMPTIME} was created to calibrate all SL BCDs, but one super dark per \texttt{RAMPTIME} \textit{per observing campaign} was created for LL BCDs. The LL super darks are a rolling average of LL CVZ observations, so the super darks have a variable level of zodiacal emission in them, making it necessary for us to determine which super dark (and therefore what zodiacal intensity) has been subtracted from each AOR. To our knowledge, the exact combinations of observations that produced the LL super darks are unavailable, so we inferred this campaign-dependent pattern by tracking the residual intensity of BCDs made from CVZ observations (data in Figure \ref{fig:cvztime}). Because the CVZ BCDs have also been processed through the \Spitzer\ pipeline and the super dark has been subtracted, their residual emission traces the history of offsets resulting from the variable zodiacal intensity. 

\begin{figure}
\centering
\includegraphics[width=0.45\textwidth]{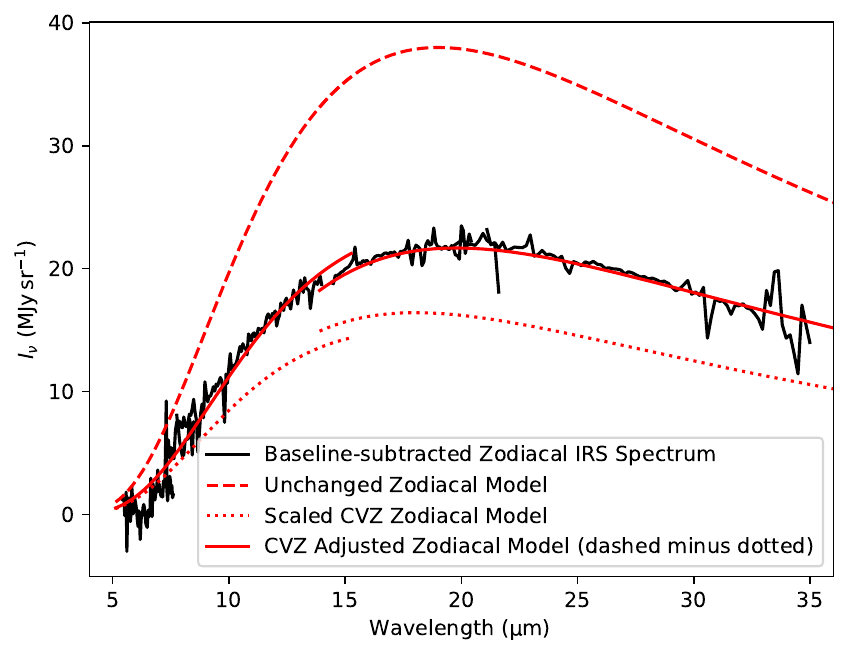}
\caption{
Example spectrum for a typical off-source observation with a decomposition of the CVZ-adjusted zodiacal emission model. The black spectrum is extracted off of the baseline frame-subtracted (see Section~\ref{sec:superdarks}) median of all BCDs within a dark AOR that only contains zodiacal emission. The dashed red line is the unchanged zodiacal emission model for the AOR ($Z_k(\lambda)$), and the dotted red line is the scaled CVZ zodiacal spectrum for the campaign that this AOR belongs to ($a(t) \hat{Z}_{\mathrm{CVZ}}(\lambda)$, (see Equation~\ref{eq:zodi}; Section~\ref{sec:zodibcds}). The disjoin at the SL--LL interface is due to the two suborders having a different set of calibration super darks in the SSC BCD pipeline. Subtracting these zodiacal models gives the solid red line, which is the final CVZ-adjusted zodiacal model spectrum for this AOR ($\mathcal{Z}_{k, ij}(\lambda)$).
}
\label{fig:zodimodel_comparison}
\end{figure}

To correct for these offsets, we created a normalized average zodiacal light spectrum at the CVZ using the ZEM ($\hat{Z}_{\mathrm{CVZ}}$), and scaled this based on the average difference between the observed (at the CVZ) and modeled zodiacal light spectrum at a certain wavelength (the centers of the SL1 and LL1 wavelength ranges, 10.95 and 28.75\,\micron, respectively) for each campaign ($a(t)$). The resulting corrected ZEM at the CVZ is shown as a function of time and at a representative wavelength in Figure~\ref{fig:cvztime}. Thus, scaling the average ZEM spectrum for the CVZ in this way mimics the average zodiacal light spectrum existing within the CVZ observations that were used to create the super darks for each campaign. These spectra were then subtracted from the 2D ZEM images for every AOR, corresponding to the observing campaign that the AOR belongs to. In Figure~\ref{fig:zodimodel_comparison}, we summarize the impact that this correction has on the spectral form of the ZEM, compared to the spectrum extracted directly from a fully-dark AOR.

\subsubsection{Baseline Frames}\label{sec:superdarks}

\begin{figure*}
\centering
\includegraphics[width=7in]{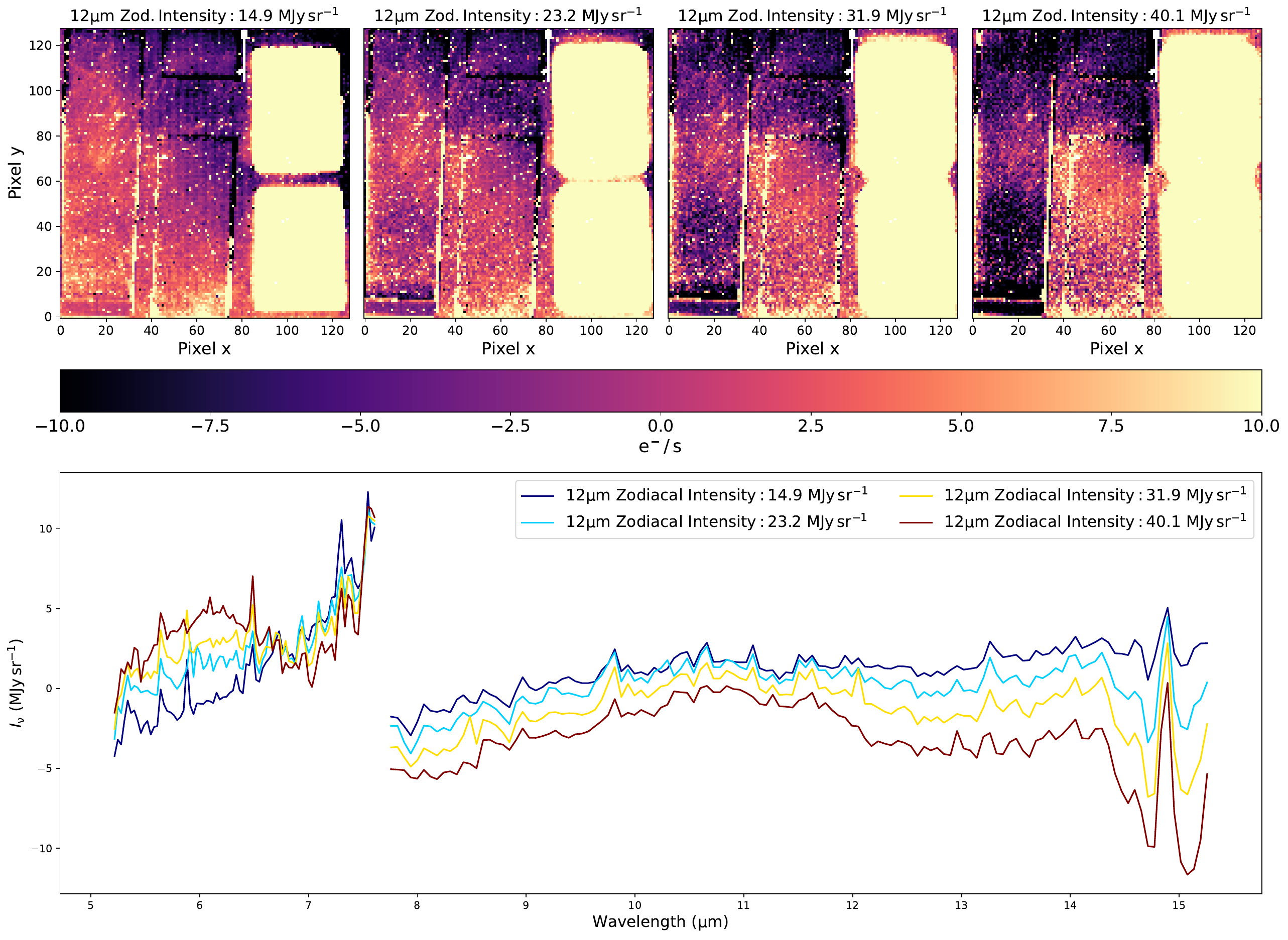}
\caption{
Top: SL binned baseline frames for a certain \texttt{RAMPTIME} (60.95\,s), with each panel in the row corresponding to a different zodiacal emission intensity. The baseline frames capture the pixel excess artifact/dark current. As the intensity increases, there are more negative (darker) pixels as the zodiacal emission model tends to overestimate. Bottom: spectra extracted from each of the above binned baseline frames. We do not show the spectra from SL3 or LL3 for clarity, because each of these overlap with other spectral suborders. The shapes of these spectra result from a combination of the artifact and systematic deviations between the observed and modeled zodiacal spectra.
}
\label{fig:superdarks}
\end{figure*}

As described above, our process for creating darks involves separately removing the zodiacal light and detector dark current signal, whereas the nominal IRS pipeline subtracts a super dark that contains a combination of both components. In the previous section we described using the ZEM to predict the zodiacal light spectrum at the date and position of each AOR and to correct that model by the portion of the zodiacal light that was already removed by the super dark (the CVZ correction). If this procedure worked exactly, BCDs pointing at blank sky should have only dark current after zodiacal light removal. However, we find that this is not the case. While there is a persistent pattern in the remaining signal that is dominated by dark current, we find that the residuals still have some dependence on zodiacal light intensity. This can be seen in Figure~\ref{fig:superdarks}, which plots the residual spectrum on the detector after zodiacal light removal.  The degree to which this spectral shape varies with zodiacal intensity may represent a combination of incorrect zodiacal emission models (both in the intensity and spectral shape) and potentially scattering of zodiacal light onto the detector, yielding a pattern that changes with the zodiacal light intensity. These two effects are difficult to separate since dark observations are never free of zodiacal emission. 

To address both of these issues simultaneously, we created ``baseline frames" binned by zodiacal light intensity for both SL and LL from a large number of dark BCDs selected from corresponding WISE photometry (see Section~\ref{sec:shards} and Section~\ref{sec:judge1}) that encapsulate the dark current and ZEM residuals. This is the second component of SIMLA backgrounds.

For pixel $ij$, a baseline frame $s$ has a value of 
\begin{equation} \label{eq:superdarks}
\begin{aligned}
s_{z^{\prime}, ij} = \mathrm{median}(d_{ij} - \mathcal{Z}_{d, ij}) \\
& \hspace*{-1.3in} \mathrm{over\ all\ }d_{ij}\mathrm{\ with} \left\{z_L<z_d<z_H\right\} \, ,
\end{aligned}
\end{equation}

\noindent where $z^{\prime}$ is the mean 12\,\micron\ intensity from the ZEM in the bin from $z_L$ to $z_H$, $d_{ij}$ is an off-source BCD with 12\,\micron\ ZEM intensity $z_d$, and $\mathcal{Z}_{d, ij}$ is the CVZ-corrected ZEM image for the AOR containing $d_{ij}$ (Equation~ \ref{eq:zodi}). Since the shape of ZEM spectra do not vary significantly, we track the zodiacal intensity of an AOR by the value at 12\,\micron\ so that it is most easily compared with WISE W3 photometry when necessary. For each baseline frame, the uncertainties from each constituent BCD are also propagated into a baseline uncertainty frame.

An AOR-tailored baseline frame, $S_{k, ij}$, was made for each AOR by linear interpolation between binned baseline frames using the 12\,\micron\ intensity from the ZEM:

\begin{equation} \label{eq:tailored_superdarks}
S_{k, ij} = I_{ij}(z_k, s_{z^{\prime}_{L},ij}, s_{z^{\prime}_{H},ij})\, ,
\end{equation}

\noindent where $z_k$ is the 12\,\micron\ intensity from the ZEM for AOR $k$, $s_{z^{\prime}_{L},ij}$ and $s_{z^{\prime}_{H},ij}$ are the nearest-in-zodiacal intensity binned baseline frames given by Equation~\ref{eq:superdarks} with $z_k>z^{\prime}_L$ and $z_k<z^{\prime}_H$, and $I_{ij}$ is the function that linearly interpolates between binned baseline frames. The uncertainties from baseline frames are also interpolated into AOR-tailored uncertainty baseline frames.

The BCDs that are used to create baseline frames are separated by module and \texttt{RAMPTIME}, and we create separate baseline frames across four bins in zodiacal intensity; four bins were found to provide the best balance between baseline frame depth and range for interpolation. For each bin, we trim pixel values using the \texttt{astropy} function \texttt{sigma\_clip}\footnote{\url{https://docs.astropy.org/en/latest/api/astropy.stats.sigma_clip.html}}, which iteratively compares and rejects pixels by comparing the same pixel between images. We use $1\sigma$ and five iterations. Finally, we median-combine the trimmed ZEM-subtracted BCDs.

Because we must bin across three parameters, some bins run out of qualifying BCDs and as a result certain baseline frames are shallow. For some bins of certain module/\texttt{RAMPTIME} combinations, no qualified BCDs are available, and only two or three bins are possible instead. The distribution of qualifying BCDs across this space is steep; the deepest baseline frames are built from $\sim 20,000$ BCDs, but the sparsest bin is built from only four. In the latter case, the baseline frame may contribute to the noise of a cube, but is still necessary to correct for artifacts and ZEM residuals.

In Figure~\ref{fig:superdarks}, we show the baseline frames for each zodiacal bin of a particular SL \texttt{RAMPTIME} as an example. In order to best mitigate the complicated interplay in pixel-space of the dark current and the ZEM, we linearly interpolate over the zodiacal intensity between binned baseline frames to create a tailored baseline frame for each AOR. An example of an interpolated AOR-specific baseline frame is shown in the top right panel of Figure~\ref{fig:dark_components}.

\subsection{Mitigating Time Variable Pixel Offsets with Shards}\label{sec:shards}

\begin{figure}
\includegraphics[width=0.45\textwidth]{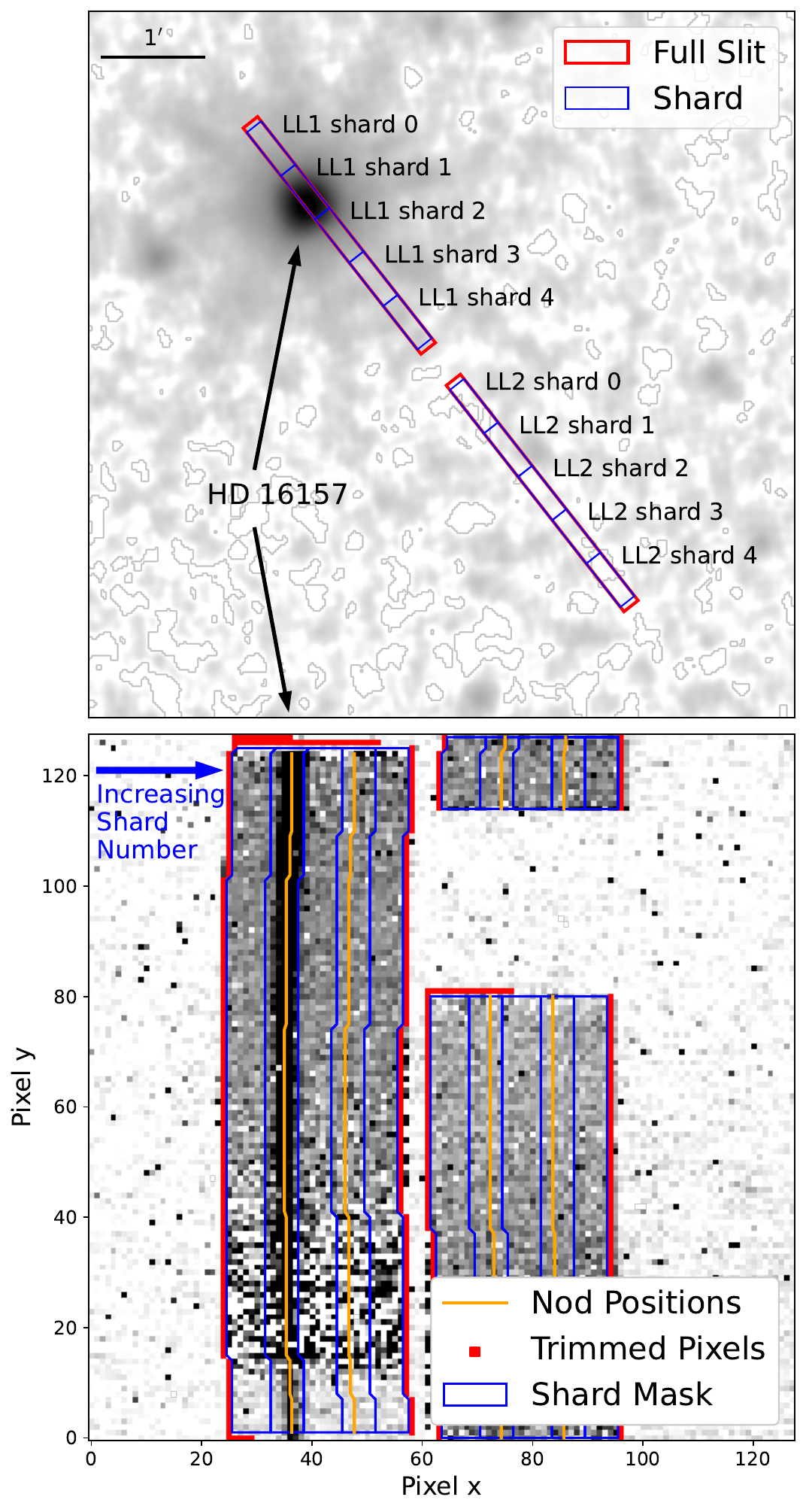}
\caption{
Top: WISE image showing an example of the LL slit positions during an observation of a star. The red rectangles show the full sky positions for the LL1 (upper) and LL2 (lower) slits. The blue rectangles illustrate the positions of the shards on the sky. Note that the shards do not span the full length because of the edge trimming. Bottom: the corresponding BCD for the observation shown above. The thick black line in the data is the spectrum of the star, which in this case is positioned on one of the nodding positions shown in orange. The blue boxes give the outlines of the masks for each shard in BCD space. Red pixels indicate that these have been trimmed.
}
\label{fig:shardmasks}
\end{figure}

The final component of SIMLA backgrounds, $D_{k, ij}$, captures the time-dependent random behavior of pixels by averaging dark IRS observations that have been baseline-subtracted and ZEM image-subtracted. See the bottom-left panel of Figure~\ref{fig:dark_components} for an example. For AOR $k$, pixel $ij$ in the dark stack $D_k$ is given by

\begin{equation} \label{eq:darkstack}
D_{k, ij} = \frac{1}{N}\sum_{n=1}^{N} \left[d_{n, ij}-(\mathcal{Z}_{d, ij}+S_{d, ij})\right]\,,
\end{equation}

\noindent where $N$ is the depth of the stack of observations (``shards," see below), $d_{n,ij}$ is a dark observation, and $\mathcal{Z}_{d, ij}$ and $S_{d, ij}$ are the CVZ-corrected ZEM image and the tailored baseline frame for $d_{n,ij}$, respectively. Below, we describe the steps necessary to assemble $D_{k, ij}$.

Portions of any BCD may contribute to a background even if other sections contain emission from bright sources; this is the case for a large number of outrigger slits, the faint edges of maps of galaxies, or similar. To take advantage of any dark section of a BCD, we subdivided each \texttt{wavesamp} area into five ``shards" along the spatial direction, as shown in Figure~\ref{fig:shardmasks}. To define the shards, we first trimmed the edges of each \texttt{wavesamp} by 8\% for SL and 4\% for LL in the spatial direction which eliminates visually permanent bad pixels that are present at the edges of the \texttt{wavesamp} on all BCDs. Because a large number of observations are pointed such that a compact source falls on one of the nod positions (indicated by orange lines in Figure~\ref{fig:shardmasks}), it is ideal to choose shard intervals such that both nod positions are centered within a shard, and the full width at half maximum (FWHM) of the IRS point-spread function (PSF) is roughly contained within one shard-width. As in the example in Figure~\ref{fig:shardmasks}, point sources will often only fall within a single shard, maximizing the number of shards that may be acceptable as components within backgrounds.

Shards appear both as sections of a BCD image (i.e.,  bottom panel of Figure~\ref{fig:shardmasks}), and as corresponding subdivided apertures on the sky (top panel of Figure~\ref{fig:shardmasks}). All-sky WISE photometry is a critical part of selecting dark regions for backgrounds as described in Section~\ref{sec:judge1}, which means that observational characteristics of the WISE images also need to be taken into account when defining shards. The FWHM of point sources within the ALLWISE W3 images is \wiseThreeFWHM\,\arcsec, so shards also need to be large enough to mitigate the effect of light lost due to the larger WISE PSF. A spacing of ten shards per channel serves as a good compromise between these considerations. This corresponds to \shardLengthSL\arcsec\ for SL shards and \shardLengthLL\arcsec\ for LL shards. In total, there are \nShards\ shards spanning SL and LL.

In the following subsections, we describe how we identified off-source shards that will be stacked in detector space to be a part of backgrounds. This process takes the form of two primary cuts: one on the WISE photometry within a shard aperture, and one on the observed spectral intensity within the shard on a BCD. We describe the reasoning and procedure behind these cuts in Sections~\ref{sec:judge1} and \ref{sec:judge2}, and our process for determining the thresholds for each cut is described in Section~\ref{sec:judge_cuts}.

\subsubsection{WISE Photometry Cut}\label{sec:judge1}

To determine if a shard contains an astronomical source or if it can be considered dark, we compare to WISE photometry at 12\um, which has a filter width that straddles the overlap between IRS SL and LL orders.

The all-sky WISE maps are co-adds over multiple orbits, meaning that any present foreground emission in an image is some combination of the varying foreground across the co-added set. Consequently, it is not practical to remove this component of WISE images with the ZEM. Instead, we applied a simple local background subtraction to each WISE image using the \texttt{photutils.Background2D}\footnote{\url{https://photutils.readthedocs.io/en/2.3.0/api/photutils.background.Background2D.html}} code.
We use this tool to create an image of the spatially varying background emission by interpolating between boxes in each quadrant of the WISE image. The backgrounds for WISE images produced this way are simple planes. We inspected a variety of WISE tiles, some dominated by point-like galaxies, some dominated by extended galaxies, and some fully covered by Milky Way emission.  We found that placing a box around each quadrant of the image and imposing a cut where the boxes are excluded if they have 90\% of pixels 3$\sigma$ above the full image median worked well by visual inspection.

After removing this background from the WISE images, we computed the sky coordinates of all shards, then obtained the average background-subtracted W3 surface brightness within these apertures. A WISE photometry cut, $C_\mathrm{W}$, is one of the checks to identify off-source shards; in Section~\ref{sec:judge_cuts}, we describe how we arrive at a particular value for this cut. Alone, this check is complicated by the fact that the WISE backgrounds are inferred locally without the context of the wider astrophysical environment; WISE images that are located at low Galactic latitudes or are pointed at bright Galactic sources will have backgrounds that include significant emission. As a result, the background-subtracted WISE photometry within these regions may appear faint even though they would contain bright sources in the IRS data. To mitigate this weakness, we imposed a second cut based on the IRS spectra themselves, described in the following section.

\subsubsection{BCD Spectrum Cut}\label{sec:judge2}

The spectra extracted from the shards on BCDs also provide information to help judge if they are pointed at blank sky. In principle, the only astrophysical emission that suitable background observations should contain is the zodiacal light. Therefore, dark shards can also be chosen by placing a cut, $C_\mathrm{s}$, on the spectral intensity of shards of BCDs once the zodiacal light and dark current are removed (via the ZEM images and baseline frames). The subtracted spectra extracted from dark shards should be emission-free across all covered wavelengths, so the second cut that we imposed for selecting background shards was on the median intensities of these spectra; see Section~\ref{sec:judge_cuts}. This cut eliminates shards with bright and extended Galactic emission that could be missed by the WISE cut due to the local background removal procedure. However, because this cut operates on spectra from individual BCDs that have not had pixel pedestals removed, the extracted shard spectra have a much lower S/N compared with the WISE photometry. Additionally, the shard spectrum cut is highly dependent on the accuracy of the combined CVZ-corrected ZEM (Section~\ref{sec:zodibcds}) and the interpolated baseline frames (Section~\ref{sec:superdarks}). This cut is therefore less sensitive than the WISE photometry cut, but the two cuts are effective for finding background shards when used in tandem.

\subsubsection{Determining the WISE Photometry and Shard Spectrum Cuts}\label{sec:judge_cuts}

\begin{figure}
\centering
\includegraphics[width=0.47\textwidth]{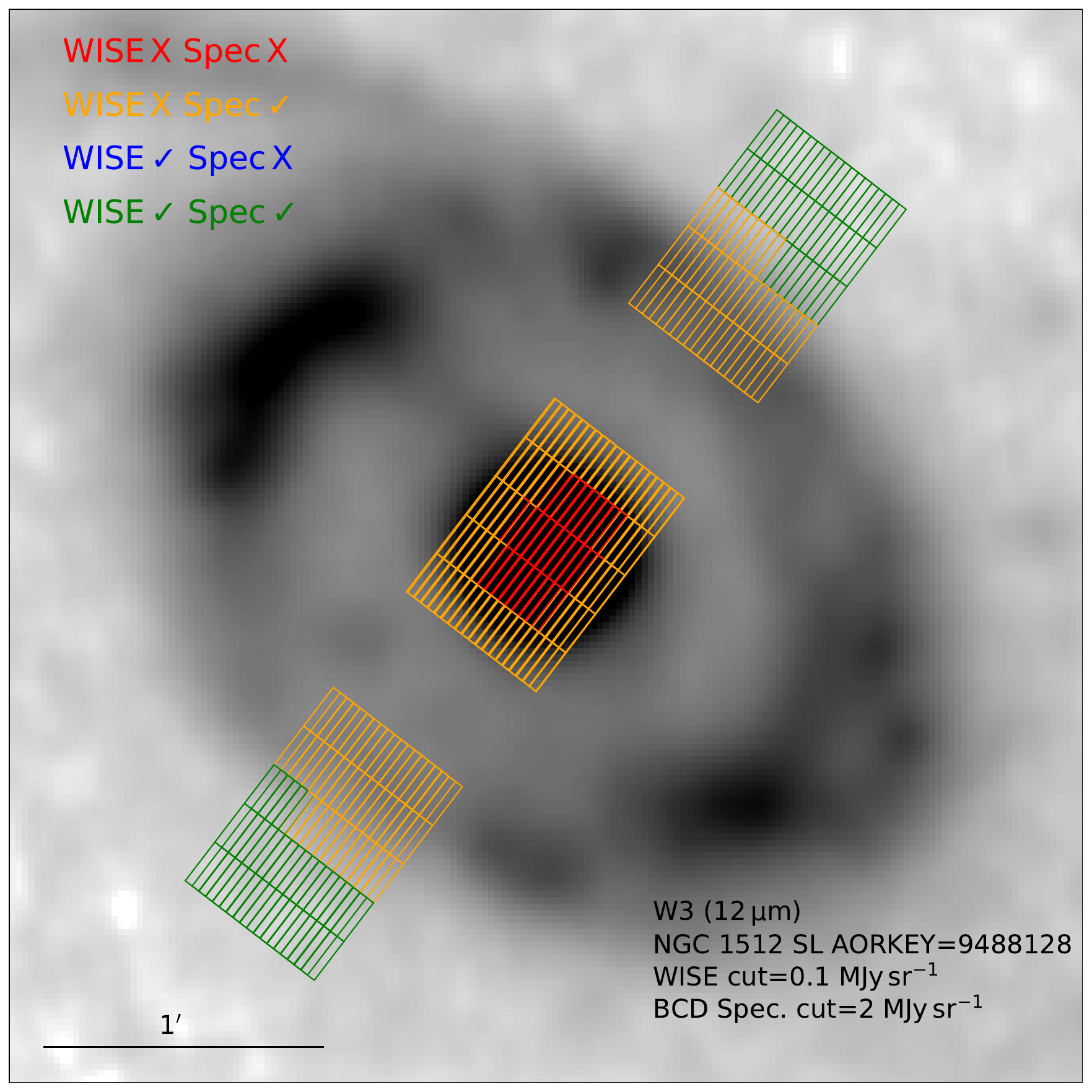}
\caption{
Shard selection diagram for a SL AOR on NGC\,1512. Colored rectangles show the sky positions of shards on a WISE image, with the colors indicating whether they passed either the WISE or BCD spectrum cut ($C_{\mathrm{W}}$, $C_{\mathrm{s}}$, respectively), both, or neither. Only green shards qualify for use as part of a background. The full color key is given in the top left, and the cut values in the lower right. In this example, there are no blue-colored cases where the WISE cut selected a shard but the BCD spectrum cut did not. Diagrams like these were used to determine the best WISE and BCD spectrum cuts for shards.
}
\label{fig:stoplight}
\end{figure}

We describe here our process to determine the particular values for $C_\mathrm{W}$ and $C_\mathrm{s}$, and we give the values used for the first SIMLA release. Updated versions of SIMLA cubes may use different values for these cuts, which will be given in the delivery documentation for each version. 

To determine the cuts to place on the WISE photometry and shard spectrum to select dark shards, we produced shard selection diagrams; see Figure~\ref{fig:stoplight} for an example. In these diagrams, the sky positions of shards are shown on top of a WISE image, with the colors of shards representing the combined results of both cuts. We visually inspected such diagrams for a range of astrophysical objects with various combinations of cut thresholds to determine the best set of cuts that selected only shards in dark regions. This analysis yielded a cut at $C_\mathrm{W}=\pm\joneval\,\mathrm{MJy\,sr^{-1}}$ for the WISE photometry, near the limit of sensitivity for the IRS. We select $C_\mathrm{s}=\pm\jtwoval\,\mathrm{MJy\,sr^{-1}}$ for the BCD spectrum cut, which is near the upper limit of offsets in dark spectra (see Section~\ref{sec:offsets}). Given the difference in ranges between $C_\mathrm{W}$ and $C_\mathrm{s}$, we find some cases where shard apertures observe Galactic ISM emission faint enough to pass $C_\mathrm{s}$, while the local WISE background subtraction allows $C_\mathrm{W}$ to be passed. Therefore, we include a final cut, $C_\mathrm{ISM}=0.5\,\mathrm{MJy\,sr^{-1}}$, where the value for each shard is estimated at the AOR-level from the ISM component of the \cite{kelsall_1998} model.

\begin{figure}
\centering
\includegraphics[width=0.45\textwidth]{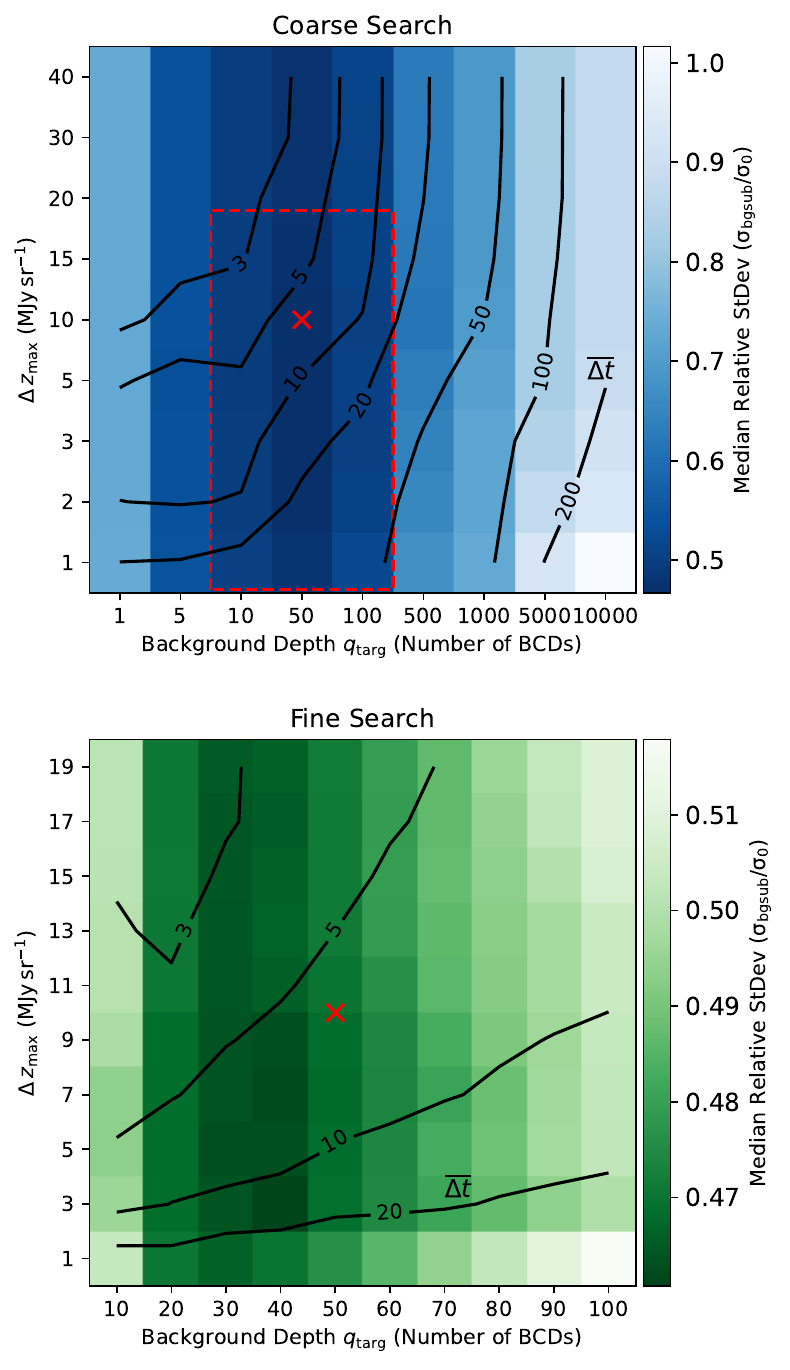}
\caption{
Results from the experiment described in Section~\ref{sec:shard_depth_cut} to find the best values for $\Delta z_{\mathrm{max}}$ and $q_{\mathrm{targ}}$. The pixel values in both panels show the median $\sigma_{\mathrm{bgsub}} / \sigma_{\mathrm{0}}$ across all test images (see text) for the combination of cuts. The top panel shows a ``coarse" parameter search, and the bottom panel shows a ``fine" search localized around the best region in the top panel, represented by the red dashed box. In both panels, the contours show the mean $\Delta t$ in days for the test backgrounds made for that set of cuts. The average ideal set of parameters is the location of the minimum $\sigma_{\mathrm{bgsub}} / \sigma_{\mathrm{0}}$, but the actual less restrictive cuts that we choose are marked by the red X.
}
\label{fig:quilts}
\end{figure}

\subsubsection{Determining the Target Shard Depth and Zodiacal Intensity Difference}\label{sec:shard_depth_cut}

Although the interpolated baseline frames mitigate systematic errors in the ZEM, we find that offsets in the background are larger and more common if shards with very different zodiacal intensities are used together (see Section~\ref{sec:offsets}). Therefore, we also impose a cut on the maximum difference between the zodiacal intensity associated with the AOR of a given cube and that of any candidate shards to be used in backgrounds ($\Delta z_{\mathrm{max}}$). The number of shards averaged together, i.e., the ``depth" of the background, must also be considered. In general, deeper backgrounds result in a cube with a higher S/N, provided that the backgrounds contain data that are observed close enough in time to the target to capture the same stochastic pixel behavior. For IRS observations, we find that using darks separated from the target by as much as several days can improve the S/N. However, once a background has reached a particular depth, additional off-source observations make the noise worse if the time difference ($\Delta t$) is too large. Because the number of available dark shards can vary widely between AORs, we aim for a target background depth, $q_{\mathrm{targ}}$, for each cube. We set up a test to determine the best values for $\Delta z_{\mathrm{max}}$ and $q_{\mathrm{targ}}$ based upon which set of cuts gives the best S/N. The results of the test are shown in Figure~\ref{fig:quilts} for both a coarse and a fine sweep of the parameter space. The basic idea is to create a large set of ``test images," each with a corresponding set of ``test backgrounds." The test backgrounds for each test image are constructed using different pairs of $\Delta z_{\mathrm{max}}$ and $q_{\mathrm{targ}}$, i.e., different locations on Figure~\ref{fig:quilts}. By quantifying how the noise within test images changes after each test background in their set is subtracted, we build up knowledge over the whole sample of what cuts produce the least noise. We assume that the results for these test images are transferrable to fully built cubes. The detailed procedure for this test is described below.

First, we identified a large group of AORs that contain at least 10 BCDs for which every shard passed $C_\mathrm{W}$, $C_\mathrm{s}$, and $C_\mathrm{ISM}$. Both the test images and test backgrounds are built from this group. Test images -- which, in principle, contain only noise -- are built for each possible AOR by subtracting the interpolated baseline frames and CVZ-adjusted ZEM images, then median-combining 10 of the BCDs in that AOR. For each test image, we produced a set of test backgrounds by averaging together similar noise-only BCDs from the large group, using BCDs that conform to a range of pairs for $\Delta z_{\mathrm{max}}$ and $q_{\mathrm{targ}}$; the zodiacal intensity of a BCD must be within $\Delta z_{\mathrm{max}}$ of that of the test image AOR, and the $q_{\mathrm{targ}}$ BCDs that are closest-in-time to the test image AOR are used. The test backgrounds always have a \texttt{RAMPTIME} and channel/suborder that match the test image. To specifically investigate the worst-case scenario where a cube has no dark observations within its own AOR, we do not allow BCDs from the same AOR as the test image to be used in the test background. The quality of each test background can then be quantified as the ratio of the noise (pixel standard deviation) within the background-subtracted test image ($\sigma_{\mathrm{bgsub}}$) to the original (i.e., not background-subtracted) level of noise in the test image ($\sigma_{\mathrm{0}}$). The results are shown in Figure~\ref{fig:quilts} for both a coarse and a fine sweep of the parameter space, to demonstrate that there is a factor of $\sim2$ difference between cuts over a large parameter space, but within the region near the minimum the dynamic range in $\sigma_{\mathrm{bgsub}}/\sigma_{\mathrm{0}}$ is small. Thus, the best set of cuts \textit{on average} is $\Delta z_{\mathrm{max}} = 3\,\mathrm{MJy\ sr^{-1}}$ and $q_{\mathrm{targ}} = 40$, but we choose $\Delta z_{\mathrm{max}} = \deltaZodiMax\,\mathrm{MJy\ sr^{-1}}$ and $q_{\mathrm{targ}} = \targDepth$ in order to be less restrictive, since the resulting noise improvement is not significantly different. In addition, we require that shards have $\Delta t < 10\ \mathrm{days}$ in order to avoid using shards that are significantly separated in time from the cube observations. This number is chosen because it is more permissive than the mean $\Delta t$ associated with the chosen set of $\Delta z_{\mathrm{max}}$ and $q_{\mathrm{targ}}$, and is near the minimum $\sigma_{\mathrm{bgsub}}/\sigma_{\mathrm{0}}$ on the bottom panel of Figure~\ref{fig:quilts}.

\subsection{Final Background}\label{sec:final_background}

To build the background for a cube, shards are collected that pass the WISE photometry, BCD spectrum, model ISM intensity, zodiacal intensity, and time difference cuts. For each AOR, we also include any (shards of) BCDs that are part of the dedicated background intended by the observer, if applicable, but are not within the same AOR. An example of this is program ID 20518, which contains three separate AORs of SL maps on M101, and a fourth AOR pointed off-galaxy for the background. In order to ensure that these ideal darks are always included in the backgrounds of their intended targets, we also include \textit{all} shards that pass the above cuts and also meet the following conditions with respect to the cube we are building a background for: 1) same program ID; 2) $\Delta t < 1\,\mathrm{day}$, and 3); an angular separation $<1\,\mathrm{degree}$.

All shards selected this way, or that meet the cuts within the cube AOR, are included in the background stack, even if there are more than $q_{\mathrm{targ}}$. If the depth of the stack after these two steps is still $<q_{\mathrm{targ}}$, we add shards from other AORs in order of increasing $\Delta t$, but always with $\Delta t < 10\ \mathrm{days}$, until $q_{\mathrm{targ}}$ is reached. Reaching the desired depth is not possible in all cases, and the depths of different shards in the background may be different. The pixels within this stack are trimmed using \texttt{sigma\_clip} ($1.5\sigma$ and three iterations), and then mean-combined to get $D_{k, ij}$. Additionally, we propagate the associated pixel $ij$ from each used BCD uncertainty frame into a combined uncertainty for $D_{k, ij}$.

Finally, pixel $ij$ of the complete background for AOR $k$, $B_{k, ij}$, has the value

\begin{equation} \label{eq:background_comps}
B_{k, ij} = \mathcal{Z}_{k, ij} + S_{k, ij} + D_{k, ij}\, ,
\end{equation}

\noindent where $\mathcal{Z}_{k, ij}$ is the CVZ-adjusted ZEM image for AOR $k$ (Equation~ \ref{eq:zodi}) and $S_{k, ij}$ is the interpolated baseline frame for $k$ (Equation~ \ref{eq:tailored_superdarks}). Similarly, the background uncertainty for $B_{k, ij}$ is the quadrature sum of the uncertainties in $S_{k, ij}$ and $D_{k, ij}$. 

\subsection{Spectral Map Assembly with \texttt{CUBISM}}\label{sec:using_CUBISM}

After the background and the background uncertainty is created for a mapping-mode AOR, they and the constituent BCDs of the AOR are input to \texttt{CUBISM} for the spectral cube assembly. \texttt{CUBISM} was designed for interactive usage, including image, cube, and spectrum viewers, and a variety of tools for validating and improving the quality of the output cube. For SIMLA, we drive \texttt{CUBISM} ``lights-out'' by directly calling underlying methods, including the automatic bad pixel flagging for global pixels. The default \texttt{CUBISM} bad pixel algorithm was used with a threshold minimum fraction of occurrence beyond 4$\sigma$ of 40\%.  The most recent \texttt{CUBISM} calibration set\footnote{\texttt{irs\_2010\_12\_22-pb-pfc-trim-omega-lhllbiasfork-slft.cal}} was used to provide extended source flux calibration, \texttt{wavesamp} area layout, the wavelength solution, and other calibration information.

\section{Quality Checks for Data Products}\label{sec:QA}

\begin{figure*}
\centering
\includegraphics[width=7in]{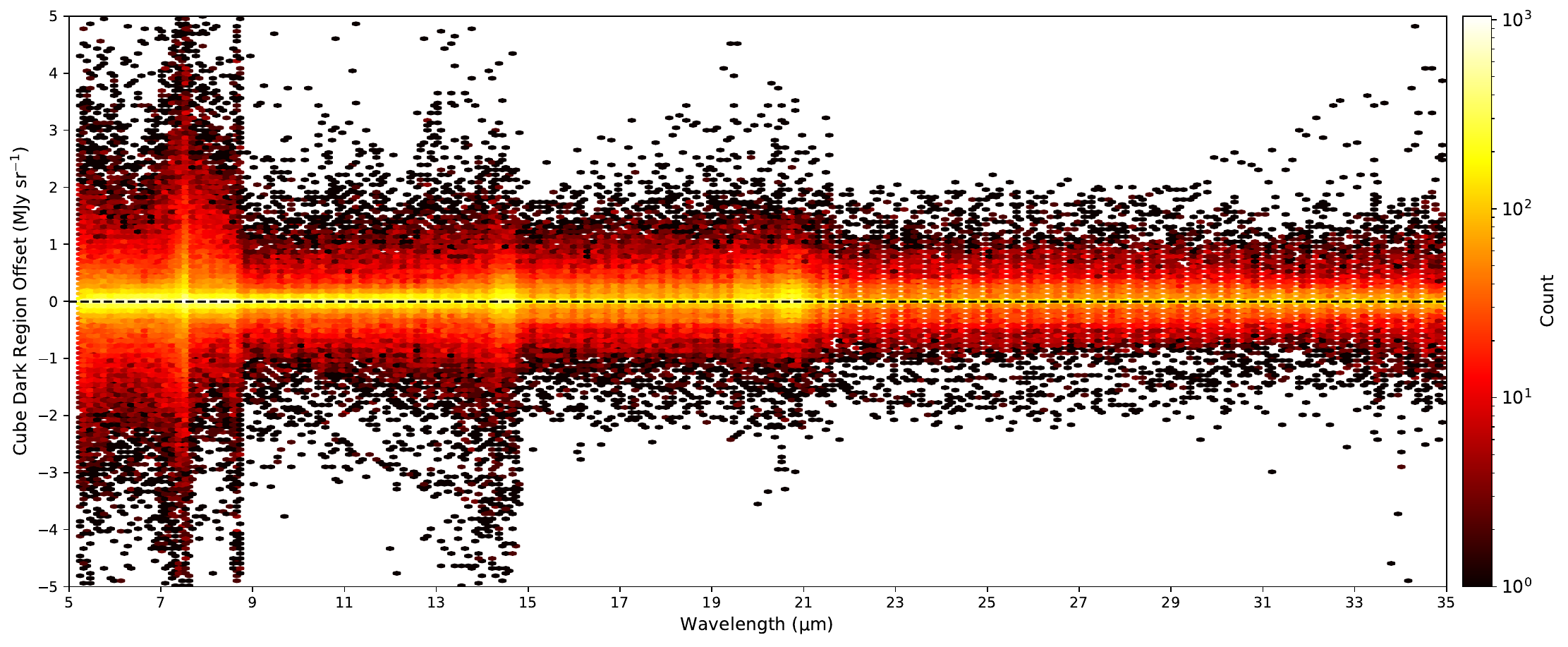}
\caption{
A two-dimensional histogram of the spectra extracted from dark regions in SIMLA cubes (see Section~\ref{sec:dark_regions}). The color bar on the right shows the density of spectral points. The vertical breaks at 7.7, 8.7, 14.5, and 21\,\micron\ align with the edges of spectral segments (SL2 and SL1, etc). By far the highest densities of SIMLA-derived surface brightnesses for dark regions are near 0\,\mjysr\ (black dashed line), as desired. 
}
\label{fig:j2_plot}
\end{figure*}

In the following subsections, we describe two tests used to validate SIMLA cubes; one test to ensure that the spectra of dark regions in cubes are indeed dark, and one test to compare the intensities of spectra from SIMLA cubes with matched WISE photometry.

\subsection{Dark Cube Regions}\label{sec:dark_regions}

Source-free sky regions provide an ideal environment for testing the quality of finished cubes because spectra from these regions should ideally have zero emission at all wavelengths. Additionally, we can evaluate the noise present in cubes without model-dependent spectral fits (Section~\ref{sec:noise}). To this end, we identified the spatial regions within SIMLA cubes that correspond to shards that passed the cuts in the background pipeline, i.e., the dark regions of SIMLA cubes; the corresponding pixels in these regions are saved as ``dark mask" files that are provided in the SIMLA release. Using the dark masks, we extract a dark spectrum for each cube where possible, and these are all shown together in Figure~\ref{fig:j2_plot}. The result shows the quality of the backgrounds; the overwhelming majority of surface brightness data points are close to 0\,\mjysr, as desired. In Section~\ref{sec:caveats}, we discuss potential reasons for any vertical offsets that remain and techniques for correcting them.

\subsection{Comparison with WISE Photometry}\label{sec:WISE_comparison}

\begin{figure}
\centering
\includegraphics[width=0.48\textwidth]{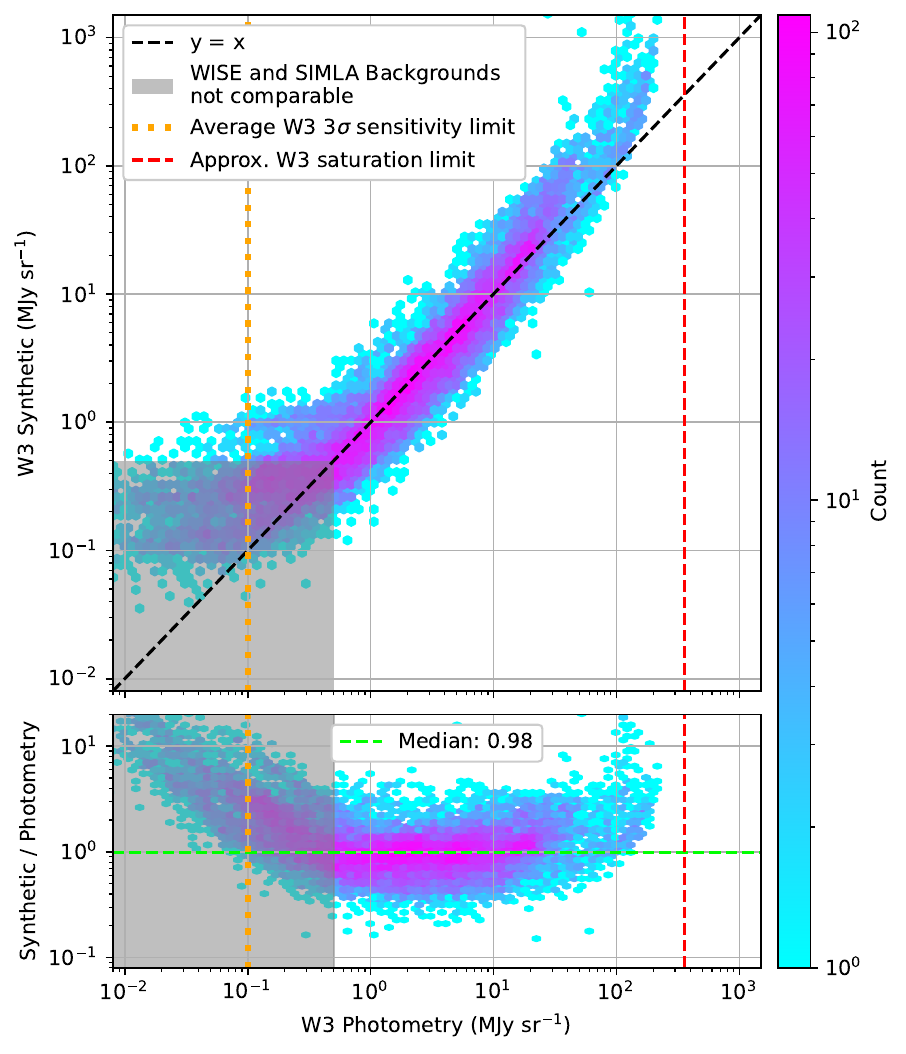}
\caption{
Top: Synthetic WISE W3 photometry derived from spectra extracted from $r=5$\,\arcsec\ apertures in SIMLA cubes, compared with matched extractions of background-subtracted WISE W3 photometry. All apertures in this figure have full IRS wavelength coverage and do not have significant Galactic ISM foreground emission. The black dashed line is the identity line. Bottom: Similar to the above, but the vertical axis indicates the ratio of synthetic to real W3 photometry. The horizontal green dashed line is the median of points with observed W3 brightness greater than $C_{\mathrm{ISM}}$. In both panels, the orange dotted line indicates the average W3 $3\sigma$ sensitivity limit ($\sim$0.1\,\mjysr), the red dashed line is the approximate saturation limit for W3 ($\sim$360\,\mjysr), and the gray region shows where the differences between the WISE and SIMLA background subtractions may be too significant for direct comparison. See Section \ref{sec:WISE_comparison}.
}
\label{fig:synthphot}
\end{figure}

Another quality test for SIMLA cubes is to check their agreement with WISE W3 photometry at fluxes above the noise floor and below the saturation limit of either instrument. For this test, spectra were extracted from SIMLA cubes and photometry was extracted from local background-subtracted WISE W3 images (see Section~\ref{sec:judge1}) using $r=5$\,\arcsec\ circular apertures in regions with full IRS wavelength coverage. This large aperture size helps mitigate light-loss effects from working with spectra across a large wavelength range. Synthetic W3 photometry is then derived from SIMLA spectra by convolving with the W3 filter curve. For this comparison, we exclude spectra from regions that do not pass $C_{\mathrm{ISM}}$.  This is because our WISE background subtraction removes all spatially smooth emission including the Galactic ISM foreground, but the background subtraction for SIMLA cubes does not (see Section \ref{sec:nonzodi_foreground}). We therefore do not expect the flux values of SIMLA cubes to agree with W3 photometry for the faint regions that do not pass this cut. The synthetic W3 photometry resulting from these spectra are compared with the observed W3 photometry in Figure~\ref{fig:synthphot}, showing excellent agreement to typically within a few percent over a large range of surface brightness. 

The agreement breaks down only at the faintest and brightest extremes of the distribution. At the faint end $\lesssim1\,\mathrm{MJy\,sr^{-1}}$, various factors that separately affect the IRS spectra and WISE photometry are likely significant enough to break the close correspondence, even if the spectra or photometry are individually interpretable. For the WISE photometry, these factors include: the WISE sensitivity limit that is pointing-dependent, but typically on the order of $0.1\,\mathrm{MJy\,sr^{-1}}$, and our WISE background subtraction, which for low-Galactic foregrounds and zodiacal intensities varies between images at the $1-2\,\mathrm{MJy\,sr^{-1}}$ level. For SIMLA IRS spectra, these factors include: small spectrophotometric offsets in cubes (see Section \ref{sec:offsets}), and noise floors that vary with \texttt{RAMPTIME}, module, and cube/background depth (see Section \ref{sec:noise}). Importantly, as mentioned, the separate background subtraction methodologies for the WISE and IRS data are not directly comparable below the value of $C_{\mathrm{ISM}}=0.5\,\mathrm{MJy\,sr^{-1}}$, indicated by the gray region in Figure \ref{fig:synthphot}, because $C_{\mathrm{ISM}}$ is a cut on the components of IRS backgrounds but not WISE backgrounds.  At such low surface brightness, it would be expected that WISE photometry would fall systematically below SIMLA synthetic photometry, as observed.

The point source saturation limit for W3 is 3.8\,mag\footnote{\url{https://irsa.ipac.caltech.edu/data/WISE/docs/release/All-Sky/expsup/sec2_2.html}}. Dividing by the solid angle of a circle with $D=12$\,\arcsec\ corresponding to the PSF FWHM of the co-added W3 images, we estimate the saturation limit in surface brightness units to be $\approx360\,\mathrm{MJy\,sr^{-1}}$. In Figure \ref{fig:synthphot}, the agreement between WISE and IRS-derived photometry at the bright end breaks down as the surface brightness approaches the WISE saturation limit \citep[IRS SL1 saturates above 5000\,$\mathrm{MJy\,sr^{-1}}$;][]{IRS}. This may be the result of nonlinearity in the WISE data at these high flux levels. However, over more than two orders of magnitude within the limits we describe, we find that the surface brightness levels in SIMLA cubes are closely matched with the WISE photometry, within a few percent, as desired.

\section{Important Notes and Caveats}\label{sec:caveats}

In this section, we describe some practical aspects of SIMLA cubes that are important to understand when interpreting SIMLA data. Additional discussion of these topics can be found in the associated data delivery document.

\subsection{Faintness Limit}\label{sec:faintness_limit}

The WISE photometry cut for SIMLA backgrounds at \joneval\,\mjysr\ imposes a surface brightness limit below which SIMLA cubes cannot be reliably interpreted. Some IRS observations are sufficiently deep to have real signal below this level, but the WISE cut allows such observations to be used as backgrounds, meaning that this real signal may be subtracted from the cube. Thus, we advise caution when interpreting spectra with an average surface brightness below \joneval\,\mjysr.

\subsection{Non-zodiacal Foreground Emission}\label{sec:nonzodi_foreground}

The desired astrophysical components that are subtracted via a background depends on the science case. In order to remain mostly impartial in this regard, the only astrophysical signal removed by SIMLA backgrounds is the zodiacal light, and there may still be unwanted signal in background-subtracted SIMLA cubes for some science cases. For example, if the science target is a galaxy near the plane of the Milky Way, the cube will likely contain a significant foreground component from the Galactic interstellar medium. In such cases, if the foreground emission is not spatially variable over the solid angle of the extragalactic target, we recommend that users subtract from the cube a 1D spectrum extracted from a suitable region in their cube that is off-target.

\subsection{Correction for Inter-order Signal}\label{sec:iocorr}

\begin{figure*}
\includegraphics[width=7in]{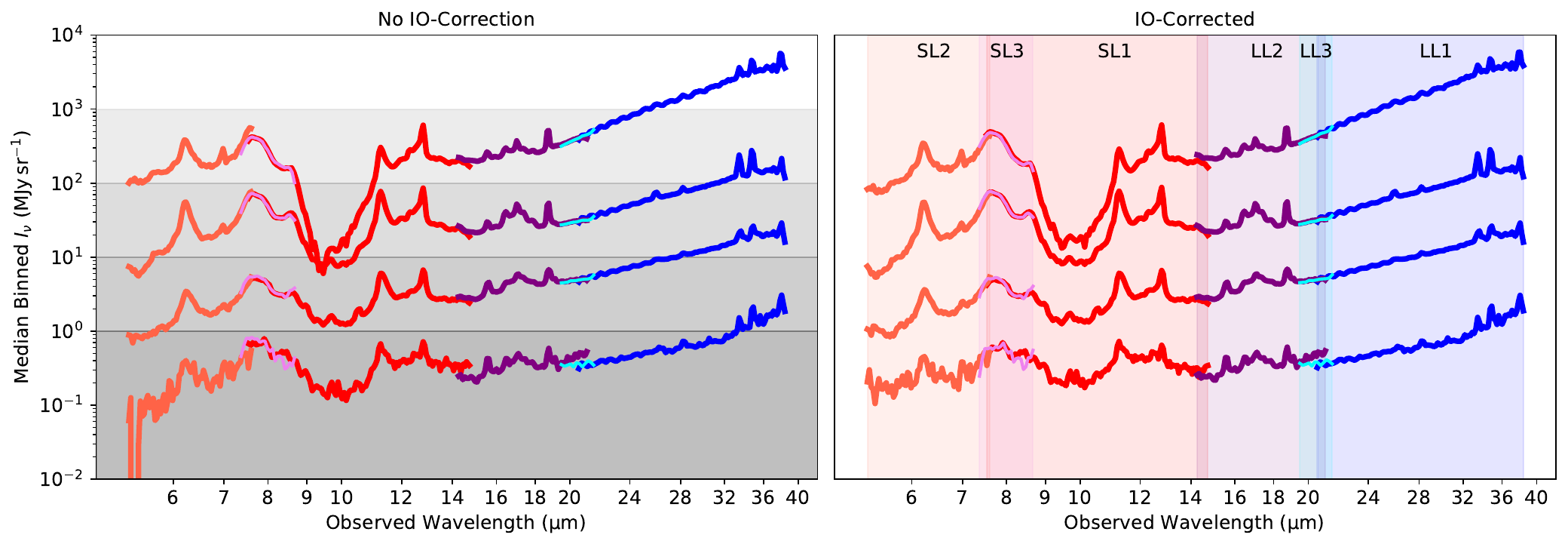}
\caption{
The difference between spectra with no correction for the inter-order (IO) signal (left) and spectra with a correction (right). The correction is only applied to SL cubes. Each spectrum is the median of extractions from SIMLA cubes binned by surface brightness; the gray regions in the left panel show the bin ranges, and the bins are the same in both panels. The spectra have not been stitched together. Colors in the spectra correspond to spectral orders (SL1, SL2, etc.), with the wavelength ranges of each shown as shaded regions in the right panel. The improvement is most noticeable at the SL2 wavelengths of the faintest bin, where the spectral shape of the IO artifact is removed. The correction also improves the ``wing" at the red end of SL2.
}
\label{fig:waterfall}
\end{figure*}

Background-subtracted IRS BCDs contain a temporally- and spatially-varying signal that can be easily seen in the inter-order (IO) region, but crosses the \texttt{wavesamp} area as well (see Figure~\ref{fig:irs_diagram}). This artifact typically manifests in spectra at a level $\mathrm{<\,1\,MJy\,sr^{-1}}$, with a similar shape as the more stable dark current (see Section~\ref{sec:AORs_and_BCDs}, Figure~\ref{fig:superdarks}). See \cite{sandstrom_2012} and \cite{starkey_thesis} for more detailed discussions and examples of procedures to correct for this. Though this signal is present in LL BCDs at a low level, we find that it is more apparent in SL cubes. Therefore, we post-process SIMLA SL cubes using the IDL code \texttt{sl\_io\_correct} from \cite{starkey_thesis}. Because the origin and shape of this correction is not well known, we choose to provide both the corrected and uncorrected cubes so that users can test whether the IO light matters for their use case. 

In Figure\,\ref{fig:waterfall}, we show a summary of the effect of this correction by showing the median of intensity-binned spectra before-and-after \texttt{sl\_io\_correct}. The correction is most noticeable in the $\mathrm{<\,1\,MJy\,sr^{-1}}$ bin, as expected, where the SL2 spectrum is flattened and there is better agreement at the SL2-SL1 interface. There are also differences for the brightest bin, in which a ``wing" at the red-end of SL2 is corrected, and the $\mathrm{\sim9.5\,\mu m}$ region is slightly flattened. In this latter case, these differences likely do not come from the IO signal, but rather from \texttt{sl\_io\_correct} correcting from stray/scattered light from the peak-up arrays that is more likely to occur near these bright sources.

\subsection{Small Offsets}\label{sec:offsets}

\begin{figure}
\includegraphics[width=0.45\textwidth]{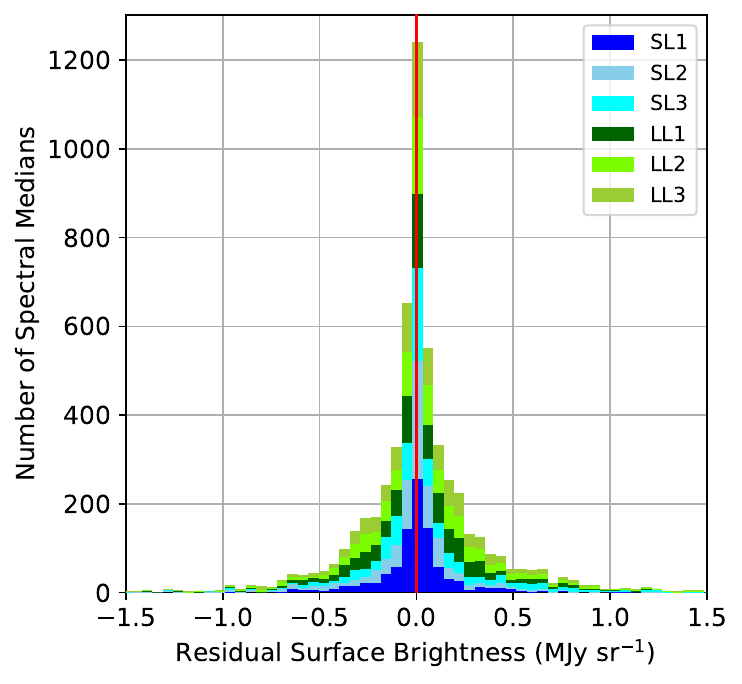}
\caption{
Similar to Figure~\ref{fig:j2_plot}, but the histogram has been collapsed along the wavelength axis to more clearly see offset values. The red line indicates the zero-line. The horizontal axis has been fixed to $\pm1.5$ for clarity. The LL segments tend to contain more offsets, likely due to the increased significance of zodiacal spectral intensity at these wavelengths (see Section~\ref{sec:offsets}).
}
\label{fig:offset_histogram}
\end{figure}

The ZEM can overestimate intensities by up to $\zodiModelPercentOverestimate\%$ when compared with the corresponding IRS observations. At the same time, there are imperfections in the CVZ correction to the ZEM, as shown by the spread of the data around the red curve in Figure~\ref{fig:cvztime}, for example. If observations selected for backgrounds of a cube have similar zodiacal intensities, then the same zodiacal emission that is subtracted from the shards is added to the background, and the ZEM makes no impact on the cube. Indeed, the same is true for the interpolated baseline frames. If, however, shards are selected with a wide distribution of zodiacal intensities, and/or these intensities are very different from the zodiacal intensity of the cube, then errors in the ZEM and CVZ correction can manifest as photometric offsets in the cube spectrum. Since these offsets are additive in nature and they do not vary spatially within a cube, we choose to demand deeper backgrounds even at the risk of offsets because the offsets can often be corrected for.

In Figure~\ref{fig:offset_histogram}, we show the median surface brightnesses of dark spectra (see Section~\ref{sec:dark_regions}), i.e., the offsets, across the entire sample. As desired, the distributions for each suborder are centered at 0\,\mjysr, indicating that significant offsets are uncommon. However, there are hundreds of cubes with offsets at the $\lesssim \offsetLevel$\,\mjysr\ level. These mostly affect LL cubes, where the zodiacal spectral intensity peaks, and thus inaccuracies in the ZEM are more significant.

These offsets can only be identified unambiguously in cubes that have source-free regions by extracting spectra there and noting a spatially-invariant deviation from zero across the spectral range of the cube. In such cases, a correction could be applied by subtracting a 1D spectrum from a dark area within the cube using the provided dark mask. However, we caution that these masks are subject to the same assumptions and limitations as the shard-selection criteria from Section~\ref{sec:shards}, and they should be used mindfully so as to not inadvertently subtract faint science targets from cubes. 

\subsection{Noise and Uncertainty Characteristics}\label{sec:noise}

\begin{figure}
    \includegraphics[width=0.45\textwidth]{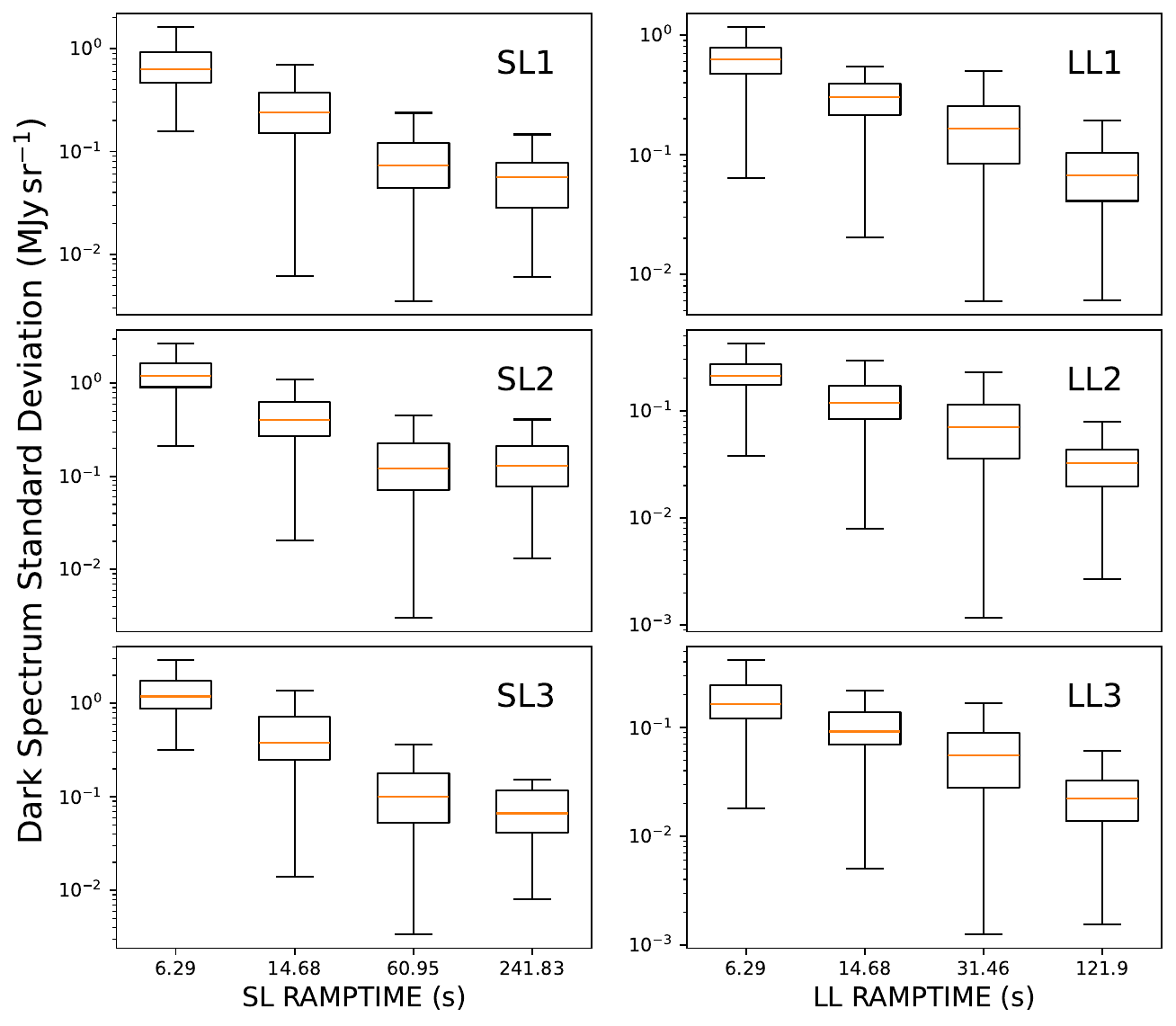}
    \caption{Summary of noise levels in the dark spectra extracted from SIMLA cubes, separated by \texttt{RAMPTIME} and suborder. The noise is quantified as the standard deviation of emission-free spectra. The box heights show the inter-quartile ranges, the orange lines represent the median values, and the whiskers indicate 1.5 times the inter-quartile range.}
\label{fig:STDV_ramptime}
\end{figure}

\begin{figure}
    \includegraphics[width=0.45\textwidth]{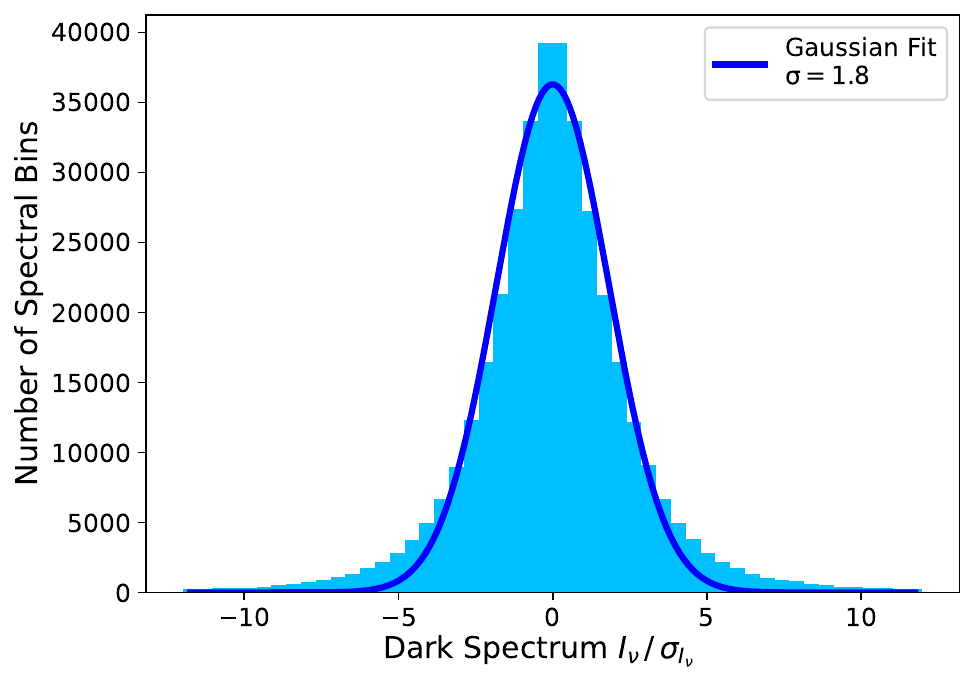}
    \caption{Distribution of surface brightness relative to the formal uncertainty for extracted dark spectra across all suborders and \texttt{RAMPTIME} values, after accounting for spectrophotometric offsets (e.g., Figure~\ref{fig:offset_histogram}). The approximate Gaussian shape with $\sigma=1.8$ of this distribution indicates close correspondence between statistical pixel and cube-level uncertainties.}
\label{fig:flux_vs_unc}
\end{figure}

The level of noise present within the cubes can be estimated using the dark region spectra described in Section~\ref{sec:dark_regions}; these spectra in principle contain no real emission, so the standard deviation of the flux across the spectrum can be attributed to noise. Note that this pixel-to-pixel noise is distinct from the surface brightness uncertainty values propagated from the uncertainty BCD images. We evaluate these dark spectra separately for the different \texttt{RAMPTIME} values and spectral orders, and show the results in Figure~\ref{fig:STDV_ramptime}. As expected, longer \texttt{RAMPTIME} settings yield lower levels of noise as these observations are deeper.

The formal uncertainties in SIMLA products are propagated from pixel-level uncertainty frames from the SSC BCD pipeline; these are primarily an estimate of BCD ramp readout noise. However, there are other uncertainties in the BCD and SIMLA pipelines that can be quantified globally across the sample. In Figure~\ref{fig:flux_vs_unc}, we show the surface brightness distribution of extracted dark spectra relative to its associated formal uncertainty, after correcting for spectrophotometric offsets by subtracting quadratic fits to the dark spectra. If the BCD ramp readout noise constituted the only noise source present in cubes, this would resemble a normal distribution, i.e., a Gaussian with $\sigma\!=\!1$. We find an approximate Gaussian shape with $\sigma\approx1.8$ from SIMLA spectra; the deviations at the peak and wings are likely attributable to imperfections in the unsupervised offset subtractions. Thus, the measured distribution indicates that the formal uncertainties are well-correlated with the offset-subtracted dark spectrum noise. The small widening by a factor of 1.8 indicates the amplitude of additional pixel-to-pixel and time-varying uncertainties apart from the ramp readout fitting, so the formal uncertainties should be interpreted as a lower limit approximation of uncertainty.

In some cases, especially for cubes with sparse pixel sampling setups, we note that there may be pixel pedestal offsets or bad pixels that persist within BCDs after the background subtraction and the global bad pixel flagging by \texttt{CUBISM}. This can result in striping artifacts in the built cube. A more robust bad pixel flagging algorithm, that takes advantage of dark observations from across the mission, is a goal for future SIMLA releases.

\section{Summary}\label{sec:summary}

\Spitzer\ spectroscopy has made enormous contributions towards our understanding of the mid-infrared universe, but until now the archive of its mapping-mode data has not reached its full potential. The \Spitzer/IRS Mapping Legacy Archive (SIMLA) provides high-quality fully-reduced spectral cubes for nearly the entire archive of \Spitzer/IRS mapping-mode observations for the first time, enabling easy access to this rich dataset. Each cube is available background-subtracted using optimized backgrounds that leverage the full IRS archive, and are made from a combination of off-source observations identified with WISE photometry and modified zodiacal emission models from \cite{kelsall_1998}. We validated the quality of the cubes by ensuring that source-free regions within them have flat spectra with a near-zero average intensity, and by checking that the synthetic photometry derived from test spectra is consistent with WISE imaging. SIMLA will soon be available at the NASA/IPAC Infrared Science Archive with a companion document for using SIMLA data products (DOI: \href{https://catcopy.ipac.caltech.edu/dois/doi.php?id=10.26131/IRSA655}{10.26131/IRSA655}).

\section*{Acknowledgments} 
We thank the anonymous referee for their
comments which improved this work. We acknowledge the critical support of the NASA/ADAP program (award ID 80NSSC21K0851), without which SIMLA could not have been achieved. This work is based [in part] on archival data obtained with the Spitzer Space Telescope, which was operated by the Jet Propulsion Laboratory, California Institute of Technology under a contract with NASA. Support for this work was provided by an award issued by JPL/Caltech. This publication makes use of data products from the Wide-field Infrared Survey Explorer, which is a joint project of the University of California, Los Angeles, and the Jet Propulsion Laboratory/California Institute of Technology, and NEOWISE, which is a project of the Jet Propulsion Laboratory/California Institute of Technology. WISE and NEOWISE are funded by the National Aeronautics and Space Administration. This research was carried out in part at the Jet Propulsion Laboratory, California Institute of Technology, under a contract with the National Aeronautics and Space Administration (80NM0018D0004).

\software{Astropy \citep{astropy:2013, astropy:2018, astropy:2022}, Matplotlib \citep{Hunter:2007}, NumPy \citep{harris2020array}, SciPy \citep{2020SciPy-NMeth}, reproject \citep{reproject}, Photutils \citep{photutils}, pandas \citep{pandas}, and Shapely \citep{shapely}}.

\bibliography{references}{}
\bibliographystyle{aasjournal}

\end{document}